\documentclass[aps,prb,showpacs,preprintnumbers,twocolumn]{revtex4-2} 
\usepackage{amsmath,amssymb}
\usepackage{bm}
\usepackage{tipa}
\usepackage{upgreek}

\usepackage{comment}
\usepackage{mathrsfs}
\usepackage{mathrsfs}
\usepackage{graphicx}
\usepackage{braket}
\usepackage{mathbbol}
\usepackage{booktabs}
\usepackage{enumitem}
\usepackage{amssymb}
\usepackage{enumitem}
\usepackage[normalem]{ulem}
\usepackage{arydshln}
\usepackage{color}
\usepackage[table,xcdraw]{xcolor}
\usepackage[colorlinks,bookmarks=true,citecolor=blue,linkcolor=red,urlcolor=blue]{hyperref}
\usepackage{cleveref}

\makeatletter
\let\old@makecaption=\@makecaption
\usepackage{subcaption}
\let\@makecaption=\old@makecaption
\makeatother

\usepackage{relsize}

\usepackage[skins]{tcolorbox}

\newtcolorbox{myframe}[2][]{%
  enhanced,colback=white,colframe=black,coltitle=black,
  sharp corners,boxrule=0.6pt,
  attach boxed title to top left={yshift=-0.3\baselineskip-0.4pt,xshift=2mm},
  boxed title style={tile,size=minimal,left=0.5mm,right=0.5mm,
    colback=white,before upper=\strut},
  title=#2,#1
}

\renewcommand{\ket}[1]{\left. \middle| #1 \right\rangle}
\renewcommand{\braket}[2]{\left \langle #1 \middle| #2 \right \rangle}
\newcommand{\braketmatrix}[3]{\left \langle #1 \middle| #2 \middle| #3 \right \rangle}

\newcommand{\MFEV}[1]{\left\langle #1 \right\rangle_\text{MF}}
\newcommand{\MFEVs}[1]{\langle #1 \rangle_\text{MF}}
\newcommand{\chern}{\mathcal{C}}

\begin{document}

\title{Paired Parton Trial States for the Superfluid-Fractional Chern Insulator Transition}

\author{Tevž Lotrič and Steven H. Simon}
\affiliation{Rudolf Peierls Centre for Theoretical Physics, Parks Road, Oxford, OX1 3PU, UK}

\begin{abstract}
We consider a model of hard-core bosons on a lattice, half-filling a Chern band such that the system has a continuous transition between a fractional Chern insulator (FCI) and  a superfluid state (SF) depending on the bandwidth to bandspacing ratio.   We construct a parton-inspired trial wavefunction ansatz for the ground states that has remarkably high overlap with exact diagonalization in both phases and throughout the phase transition.  Our ansatz is stable to adding some bosonic interactions beyond the on-site hard core constraint. We confirm that the transition is well described by a projective translation symmetry-protected multiple parton band gap closure, as has been previously predicted.  However, unlike prior work, we find that our wavefunctions require anomalous (BCS-like) parton correlations  to describe the phase transition and SF phase accurately. 

\end{abstract}

\maketitle

Much of our understanding of fractional quantum Hall (FQH) physics
comes from 
wavefunctions which have very high overlap with exact 
diagonalization
results.       For fractional Chern insulators (FCIs)~\cite{SidFCIReview,BergholtzReview,BergholtzReview2,Park2023,FCIPhysRevX.13.031037,Lu2024}, the lattice analog of FQH states, similarly good trial wavefunctions are only known to exist in a few special cases that can be mapped to a FQH problem --- in particular for so-called ideal~\cite{RahulRoyIdealBand} or vortexable~\cite{LedwithVortexable,HigherVortexable} bands with zero band width~\cite{KapitMueller}.    In this paper, we investigate trial wavefunctions for FCIs suggested by the parton construction for the bosonic $\nu=\frac12$ state, demonstrating that they can have very large overlap with the exact ground state --- even when the underlying band has significant band width.   We show the value of these wavefunctions by using them to examine a particular phase transition in detail.

The possibility of a continuous transition between a topologically-ordered phase and more conventional phases is an interesting aspect of FCI physics which is largely absent for FQH. Understanding conditions under which such transitions can occur could enable easier preparation of many-particle cold-atom FCI states \cite{pollman_state_prep_PhysRevB.96.165107,PhysRevB.103.085122,04_meng_numerics,05_barkMcGreevyTheory,demler_coldatom_FCI_PhysRevLett.94.086803,17_PhysRevLett.115.026802}. Furthermore, there is fundamental interest in these transitions as they are examples of transitions beyond the Landau-Ginzburg parardigm. Certain transitions out of topological phases are also predicted to show emergent symmetries absent at the microscopic level \cite{15_Senthil_theory,53_zaletel_vishwanath_PhysRevX.8.031015,18_detailed_transition_with_dualities}. Perhaps the clearest example of such a transition is that between the $\nu=\frac12$ bosonic FCI and a conventional superfluid (SF) which we study here.
After an initial proposal of how a continuous FCI-SF transition may be described within the parton picture \cite{05_barkMcGreevyTheory}, there has been some recent work refining the theory \cite{15_Senthil_theory} and also a recent numerical paper providing evidence for a continuous transition using DMRG~\cite{04_meng_numerics}.   Despite this, it is not clear whether the transition mechanism of Refs.~\cite{05_barkMcGreevyTheory,15_Senthil_theory} is realized in the numerics. Furthermore, the parton picture of this transition relies on a sequence of non-trivial assumptions about the nature of the parton mean-field ansatz, some of which we point out later. It is not immediately obvious if the parton picture leads to a good description of either phase, much less the transition itself. Inspired by the parton picture, we construct a variational wavefunction ansatz which we demonstrate to have an excellent overlap with the exact ground state over the entire transition. This answers all of the above concerns: the high overlap gives a solid footing to the parton-based approach, while the parton interpretation of our wavefunction largely confirms that the transition is of the type discussed by Refs.~\cite{05_barkMcGreevyTheory,15_Senthil_theory}.   However, in contrast to previous work on partons, we find that we need to introduce anomalous parton correlators (analogous to BCS theory) to describe the phase transition and SF phase accurately. 

\textit{Model.---} A simple system that can display this transition is a lattice of hard-core, but otherwise non-interacting bosons half-filling a band with Chern number $\chern=1$. For special lattice models with a nearly-flat Chern-band limit, we expect to see a FCI phase, while in a dispersive Chern band we generically expect a superfluid phase. The presence of the FCI phase is well-established for example for the \textit{checkerboard} and the \textit{honeycomb} lattice models \cite{04_meng_numerics,39_PhysRevLett.107.146803,41_PhysRevLett.106.236804,42_PhysRevLett.106.236803}. We begin by focusing on these models, so Hamiltonians of the form
\begin{equation} \label{eq:bose_ham}
  H^\text{B}=\sum_{\mathbf{x},\mathbf{y}}t({\mathbf{x},\mathbf{y}})b^\dagger(\mathbf{x})b(\mathbf{y}),
\end{equation}
together with the hard-core constraint $b(\mathbf{x})^2=0$. Here $\mathbf{x}$ labels lattice sites and in what follows, the hoppings $t(\mathbf{x},\mathbf{y})=t^*(\mathbf{y},\mathbf{x})$ are assumed to have lattice translation symmetry, with details for the two models discussed in Sect.~\ref{sup:hoppings} of the Supplement~\cite{Supplement}.

\textit{The parton construction.---} The FCI state may be described using the parton construction \cite{05_barkMcGreevyTheory,wen_book,11_Wen_1999}, where we write the bosonic operator as
\begin{equation} \label{eq:parton_decomposition}
    b(\mathbf{r})=f_1(\mathbf{r})f_2(\mathbf{r})
\end{equation}
with $f_{1,2}$ fermionic operators. This naturally enforces the hard-core constraint $b^2(\mathbf{r})=0$, but it doubles the Hilbert space dimension per site. We should only concern ourselves with the physical subspace where $\hat{n}_1(\mathbf{r})=\hat{n}_2(\mathbf{r})=\hat{n}_\text{B}(\mathbf{r})$ with $\hat{n}_{1,2}(\mathbf{r})=f_{1,2}^\dagger(\mathbf{r})f_{1,2}(\mathbf{r})$ and $\hat{n}_\text{B}(\mathbf{r})=b^\dagger(\mathbf{r})b(\mathbf{r})$. In a field-theoretical description, this constraint is enforced by coupling the partons to an $SU(2)$ gauge field. The advantage of partonizing is that the parton mean-field turns out to describe certain phases, including the FCI well as it allows us to expand around a non-trivial mean-field point even when $\langle b \rangle=0$ \cite{wen_book}. In particular, using the parton decomposition Eq.~\ref{eq:parton_decomposition} in Eq.~\ref{eq:bose_ham} and applying the mean-field (MF) approximation, we find up to a constant
\begin{equation} \label{eq:simple_mf_expansion}
\begin{split}
    H^\text{MF}=&\sum_{\mathbf{x},\mathbf{y}}t({\mathbf{x},\mathbf{y}})\left[\MFEV{f_1^\dagger(\mathbf{x})f_1(\mathbf{y})}f_2^\dagger(\mathbf{x}) f_2(\mathbf{y})\right.\\
    &\left.+\MFEV{f_2^\dagger(\mathbf{x}) f_2(\mathbf{y})}f_1^\dagger(\mathbf{x})f_1(\mathbf{y})\right]
  \end{split}
\end{equation}
where we assumed for simpliciy that the two partons are independent at the MF level. This assumption is relaxed in Eq.~\ref{eq:general_MF_ham}. 
To understand Eq.~\ref{eq:simple_mf_expansion}, which is a free Hamiltonian, we need to address a subtlety relating to translational symmetry. We impose that $H^\text{B}$ is invariant under two bosonic translation operators $T_{1,2}^\text{B}$ which correspond to translations by the primitive lattice vectors and obey $T^\text{B}_1 T^\text{B}_2 = T^\text{B}_2 T^\text{B}_1$. We may similarly define translational operators in the parton language $T_{1,2}^\text{P}$, but the partons only need to fall into a \textit{projective} representation of this symmetry, in particular we may have an ansatz where $T^\text{P}_1 T^\text{P}_2=-T^\text{P}_2T^\text{P}_1$ \cite{wen_book,25_McGreevy_wavefunction_PhysRevB.85.125105}. A physical intuition for this comes from Ref.~\cite{05_barkMcGreevyTheory} where a Landau level is considered instead of a lattice $\chern=1$ band. In that case, the unit-charge boson sees $2\pi$ flux per plaquette, but Eq.~\ref{eq:parton_decomposition} dictates that the two partons share the boson's charge. Assuming they each have half-unit charge, they each see $\pi$-flux, leading to the above magnetic translation relations for $T^\text{P}_{1,2}$, effectively doubling the unit-cell. The fact that both partons see equal charge is a non-trivial assumption about the nature of the parton MF state.   The success of our approach (below) 
confirms such an assumption is sensible, but we should note that it is far from guaranteed --- for example for the same Hamiltonian, but for
filling $\nu=\frac23$, the two partons would see \textit{different} charges and we would get a tripling of the unit cell instead. For $\nu=\frac12$ we checked that other extensions of the unit cell do not lead to an improvement in the variational wavefunction. The magnetic translation relations for $T^\text{P}_{1,2}$ mean that the partons see a doubled unit cell which we take to be generated by the commuting $(T^\text{P}_{1})^2$ and $T^\text{P}_{2}$ \cite{05_barkMcGreevyTheory,25_McGreevy_wavefunction_PhysRevB.85.125105}. Since the initial problem was at half-filling, each flavor of partons fills a band at the MF level. The Chern numbers of these bands determine the phase of the bosonic system. If $t(\mathbf{x},\mathbf{y})$ and the expectation value in Eq.~\ref{eq:simple_mf_expansion} are such that the partons have Chern numbers $\chern^\text{P}_1=\chern^\text{P}_2=1$, it can be shown that the system is in a FCI phase. The phase $\chern^\text{P}_1=1$, $\chern^\text{P}_2=0$ is a Mott insulator and $\chern^\text{P}_1=1$, $\chern^\text{P}_2=-1$ is a superfluid \cite{05_barkMcGreevyTheory}. While the argument in Ref.~\cite{05_barkMcGreevyTheory} is field-theoretical, much of this may be understood in terms of wavefunctions.

Let $\ket{\Omega_N}$ be the ground state of Eq.~\ref{eq:simple_mf_expansion} with $N$ partons of each flavor. Given the non-interacting Hamiltonian, $\ket{\Omega_N}$ is a Slater determinant for each parton. Together with Eq.~\ref{eq:parton_decomposition} it suggests a bosonic wavefunction,
\begin{equation} \label{eq:bose_wf_form_mf}
  \phi(\mathbf{r}_1,\ldots,\mathbf{r}_N)=\braketmatrix{\emptyset}{b(\mathbf{r}_1)\ldots b(\mathbf{r}_N)}{\Omega_N}.
\end{equation}
If the parton flavors in $\ket{\Omega_N}$ are independent, this implies that Eq.~\ref{eq:bose_wf_form_mf} is the product of two Slater determinants, $\phi(\{\mathbf{r}_j\})=\det\left[ \varphi_k^{(1)}(\mathbf{r}_j)\right] \det\left[ \varphi_l^{(2)}(\mathbf{r}_j)\right]$. 
This is not trivial however since both determinants have the same positions as arguments. In the simple case of each parton filling the LLL, each determinant may be written (neglecting Gaussian factors) as $\prod_{i<j}(z_i-z_j)$ and so Eq.~\ref{eq:bose_wf_form_mf} gives $\phi =  \prod_{i<j}(z_i-z_j)^2$, the bosonic $\nu=\frac12$ Laughlin wavefunction. Analogously, both partons filling $\chern^\text{P}=1$ bands implies a $\nu=\frac12$ FCI. The case $\chern^\text{P}_1=-\chern^\text{P}_2$ can be understood by looking at the limit $\varphi^{(2)}_l(\mathbf{r}) = \left(\varphi^{(1)}_l(\mathbf{r})\right)^*$, where it is easy to show that the resulting $\phi$ has ODLRO and represents a superfluid.

Armed with an understanding of the parton description of both phases, Ref.~\cite{05_barkMcGreevyTheory} argued that if the effective Hamiltonian Eq.~\ref{eq:simple_mf_expansion} is varied so that parton 1 remains gapped, while parton 2 undergoes a double Dirac-cone gap closure sending $\chern^\text{P}_2 = 1\rightarrow -1$, we would have a continuous transition between the FCI $\chern^\text{P}_2 = 1$ and the SF $\chern^\text{P}_2 = -1$ states. It was later pointed out that this unusual double gap closure is protected by the projective action of the translation operators on the partons \cite{15_Senthil_theory,18_detailed_transition_with_dualities,53_zaletel_vishwanath_PhysRevX.8.031015}. The projective translations guarantee a doubling in the spectrum as all representations of $T^\text{P}_1T^\text{P}_2=-T^\text{P}_2T^\text{P}_1$ are at least two-dimensional. Denoting by $\ket{\psi_\mathbf{k}}$ with $\mathbf{k}=(k_x,k_y)$ a one-particle state with eigenvalues $e^{ik_x}$ and $e^{ik_y}$ under $(T^\text{P}_1)^2$ and $T^\text{P}_2$, we find that $T^\text{P}_1 \ket{\psi_\mathbf{k}}=\ket{\psi_\mathbf{\tilde{k}}}$ with $\mathbf{\tilde{k}}=(k_x,k_y+\pi)$ is degenerate to $\ket{\psi_\mathbf{k}}$. So if a gap closure occurs at $(q_x,q_y)$, it must also occur at $(q_x,q_y+\pi)$, implying that the parton Chern numbers jump by two along transitions unless translational symmetry is broken (which is the case when transitioning into the Mott phase). The critical theory of the transition is then two gapless Dirac fermions, coupled to an emergent  Chern-Simons $U(1)$ gauge field (the initial $SU(2)$ introduced by the partons is assumed to be Higgsed to $U(1)$) \cite{05_barkMcGreevyTheory}.

We again point out that it is far from obvious that this is a physically reasonable scenario. While cylinder-DMRG has indeed shown evidence for a continuous transition \cite{04_meng_numerics} and it may also be used to probe the predicted $SO(3)$ CDW fluctuation symmetry at the transition \cite{15_Senthil_theory,53_zaletel_vishwanath_PhysRevX.8.031015,wang2025emergentqed3bosoniclaughlin}, we believe a more direct evaluation of the validity of the parton picture of this transition is necessary. We address these concerns by developing a variational wavefunction ansatz based on the parton picture and demonstrating superb agreement with other numerical methods. We provide evidence that the transition may indeed be understood as a jump in the parton Chern numbers $\chern^\text{P}$ in a situation where a double gap closure is protected by the projective action of translations.

\begin{figure}
    \centering
    \includegraphics[width=3.5in]{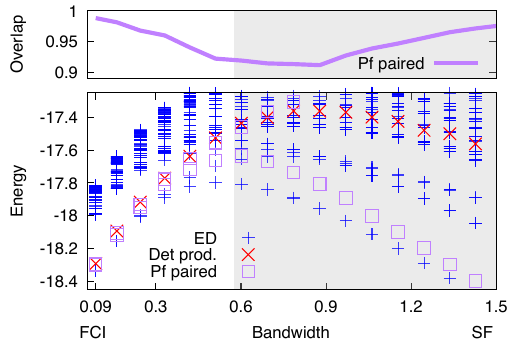}
    \caption{Comparison of both paired (Pf) and uncorrelated (det prod.) parton wavefunction ansatz from VMC against exact diagonalization (ED) in a 9-boson system (total Hilbert space dimension $10^8$) as we move from a FCI phase to SF (shaded) by increasing bandwidth.   (Top) Overlap of exact ground states with our paired ansatz for the $\mathbf{Q}=(0,0)$ state which is the lowest energy branch.  (Bottom) energies of trial states versus ED.   Both trial ground states are shown.  Deep into the FCI phase these two are near degenerate, and the Det prod. wavefunction is also accurate.   However, going into the SF phase, the two paired wavefunctions split with the $\mathbf{Q}=(\pi,0)$ and the Det. Prod. wavefunctions going up in energy. Note that the gap seen on the SF side is a finite-size effect, while the FCI gap remains finite for large systems.   The horizontal axis does not extend down to zero bandwidth as we would need longer ranged hopping to achieve this.
    }
    \label{fig:checkerboard_ed_comparison}
\end{figure}

We might first try to optimize over all possible ansatze of the determinant-product form described below Eq.~\ref{eq:bose_wf_form_mf} as this general form can capture both sides of the transition. While such a procedure sees a transition, we find it to be a relatively poor description of the superfluid phase when comparing to exact diagonalization, as we show in Fig.~\ref{fig:checkerboard_ed_comparison}. The reason for this may be traced back to the fact that the partons are assumed independent in the MF state, so $\MFEV{b}=\MFEV{f_1 f_2}=0$ in that state. Such MF-independent partons can display a superfluid phase \cite{05_barkMcGreevyTheory}, but the resulting SF wavefunction is not accurate. In Sect.~\ref{sup:sf_field_theroy} of \cite{Supplement}, we develop a field-theoretical picture showing that parton states with a nonzero $\MFEVs{f_1f_2}$ describe the same SF phase and that the MF saddle-point generally has a nonzero $\MFEVs{f_1f_2}$. This motivates considering paired parton trial states.


We note here the distinction between the MF and true expectation values --- for an operator $\mathcal{O}$ and a MF state $\ket{\Omega}$, we have $\MFEV{\mathcal{O}}=\braketmatrix{\Omega}{\mathcal{O}}{\Omega}$ while the true expectation value in the state $\ket{\phi}\propto P\ket{\Omega}$ projected onto the physical subspace (eg. Eq.~\ref{eq:bose_wf_form_mf}) is different. In operator language, $P=\prod_\mathbf{r} \left[\hat{n}_1(\mathbf{r})\hat{n}_2(\mathbf{r})+(1-\hat{n}_1(\mathbf{r}))(1-\hat{n}_2(\mathbf{r}))\right]$. Any valid bosonic operator (including $H^\text{B}$) must commute with the projector $P$ \footnote{This translates to requiring $\mathcal{O}$ to be a singlet w.r.t. the $SU(2)$ gauge field in the field-theoretical approach.}. This, together with $P^2=P$, implies for the projected expectations $\braketmatrix{\phi}{\mathcal{O}}{\phi}=\langle\mathcal{O}\rangle=\braketmatrix{\Omega}{\mathcal{O} P}{\Omega}/\braketmatrix{\Omega}{P}{\Omega}=\frac{\MFEV{\mathcal{O}P}}{\MFEV{P}}$. The MF expectations evaluate operators on both the physical and unphysical subspaces --- still, MF quantities such as Chern numbers are useful to interpret the projected state.

\textit{Paired parton states.---} We choose to work with the most general mean-field ansatz. For $K$ flavors of partons ($K=2$ in our case), we can write the most general MF Hamiltonian as
\begin{equation} \label{eq:general_MF_ham}
\begin{split}
    H^\text{MF} =\sum_{\mathbf{x},\mathbf{y}}\sum_{\alpha,\beta=1}^K &t^\text{MF}_{\alpha\beta}(\mathbf{x},\mathbf{y}) f_\alpha^\dagger(\mathbf{x})f_\beta(\mathbf{y})\\ +& \Delta^\text{MF}_{\alpha\beta}(\mathbf{x},\mathbf{y}) f_\alpha(\mathbf{x})f_\beta(\mathbf{y}) + \text{h.c.}
\end{split}
\end{equation}
for some $t^\text{MF},\Delta^\text{MF}$. Considering possible parton contractions in Eq.§~\ref{eq:bose_ham}, it may seem that many of the terms in Eq.~\ref{eq:general_MF_ham} must be zero. But the operator we should really consider is not $H^\text{B}$, but $H^\text{B}P$. This will lead to more contractions, possibly over larger distances. While this \textit{allows} for any fermionic bilinear to have a nonzero MF expectation, depending on $H^\text{B}$, the best variational wavefunctions may occur when some bilinears have $\MFEVs{\cdot}=0$. Instead of explicitly evaluating the coefficients in Eq.~\ref{eq:general_MF_ham}, we note that the ground state, projected to a fixed number of bosons, will always be a Pfaffian. Say we have $M$ partons ($M=2N$ for $N$ bosons) at positions $\mathbf{r}_j$ and of flavors $\alpha_j$ for $j=1\ldots M$. Then, the ground state is given by the Pfaffian form \cite{46_Read_Green}
\begin{equation} \label{eq:mf_Pfaffian_form}
\Omega^\text{MF}\left(\{(\mathbf{r}_j,\alpha_j)\}_{j=1\ldots M}\right)=\text{Pf}\left[g_{\alpha_j\alpha_k}(\mathbf{r}_j,\mathbf{r}_k)\right]_{j,k=1\ldots M}
\end{equation} 
for some function $g_{\alpha\beta}(\mathbf{x},\mathbf{y})=-g_{\beta\alpha}(\mathbf{y},\mathbf{x})$. Due to the possible presence of terms such as $f_1^\dagger f_2$ in Eq.~\ref{eq:general_MF_ham}, the partons may change flavor. We can think of flavor as an additional site index --- each \textit{physical} site $\mathbf{r}$ splits into $K$ sub-sites indexed by $(\mathbf{r},\alpha)$ for $\alpha=1\ldots K$. The parton `flavor' then simply specifies at which of these sub-sites the parton resides. In this language, the physical subspace corresponds to states where each \textit{physical} site either has all $K$ of its sub-sites occupied or all of them empty. In our $K=2$ case, a state with $N$ bosons at positions $\mathbf{r}_1,\ldots,\mathbf{r}_N$  is described by filling both sub-sites at each of those sites in Eq.~\ref{eq:mf_Pfaffian_form}, giving the bosonic wavefunction
\begin{equation} \label{eq:projected_bose_WF}
  \phi(\{\mathbf{r}_j\}_{j=1\ldots N})=\Omega^\text{MF}(\{(\mathbf{r}_j,1),(\mathbf{r}_j,2)\}_{j=1\ldots N}).
\end{equation}
This obeys both bosonic statistics and the hard-core constraint. It is equivalent to using a paired MF state $|\tilde\Omega\rangle$ in Eq.~\ref{eq:bose_wf_form_mf}. Our method in what follows is to take the form of Eq.~\ref{eq:projected_bose_WF} as our wavefunction ansatz, using the values of the pairing function $g_{\alpha\beta}(\mathbf{x},\mathbf{y})$ as our variational parameters. As $\phi$ is an ansatz for the full bosonic wavefunction, the optimization scheme will be to directly minimize $E[\phi]=\frac{\braketmatrix{\phi}{H^\text{B}}{\phi}}{\braket{\phi}{\phi}}$, bypassing the need to evaluate $H^\text{MF}$ in Eq.~\ref{eq:general_MF_ham}. The evaluation of $E[\phi]$ is however not analytically tractable, forcing us into variational Monte-Carlo (VMC) sampling. Details of the sampling and optimization are shown in the Supplement~\cite{Supplement}, Sect.~\ref{sup:alg_trickery}. Our VMC procedure is different from any mean-field scheme (e.g. solving Eq.~\ref{eq:simple_mf_expansion} self-consistently) as it optimizes $\langle H^\text{B}\rangle=\MFEV{H^\text{B}P}/\MFEV{P}$ and not $\MFEV{H^\text{B}}$. 

We require that the function $g_{\alpha\beta}(\mathbf{x},\mathbf{y})$ in Eq.~\ref{eq:mf_Pfaffian_form} obeys the translational symmetry mandated by the larger, partonic unit cell (so $(T^\text{P}_1)^2$ and $T^\text{P}_2$). We do not enforce any action of $T^\text{P}_1$ itself on $g_{\alpha\beta}(\mathbf{x},\mathbf{y})$ --- this allows the total bosonic-unit-cell momentum to be $\mathbf{Q}=(0,0)$ and $\mathbf{Q}=(\pi,0)$. By considering $\phi(\{\mathbf{r}\})\pm\phi(\{\mathbf{\tilde{r}}\})$ with $\{\mathbf{\tilde{r}}\}$ having all particles translated by $T_1^\text{B}$ relative to $\{\mathbf{r}\}$, we can construct an ansatz for each of the above $\mathbf{Q}$. For system dimensions where the two topologically degenerate states in the FCI phase occur at different momenta, this will allow us to generate both states. 

\textit{Results and comparison to exact diagonalization.---} Similarly to Ref.~\cite{04_meng_numerics}, we start with the FCI in the flatband limit of the checkerboard model and vary the hopping parameters (details in Sect.~\ref{sup:hoppings} of \cite{Supplement}) so that the dispersion develops a finite bandwidth with a minimum at the $\Gamma$-point, $\mathbf{k}=(0,0)$ in the boson Brillouin zone. The magnitude of hopping parameters is $\sim\max(|t|)=1$ with the bandgap at the same scale. In Fig.~\ref{fig:checkerboard_ed_comparison}, we compare our best VMC result to ED for a system of $9$ bosons on a $6\times3$ torus. In the FCI phase, the two topologically degenerate ground states occur at total momenta $\mathbf{Q}=(0,0)$ and $\mathbf{Q}=(\pi,0)$, while the SF phase has a unique GS at $\mathbf{Q}=(0,0)$. Our VMC procedure can find both of the topologically degenerate states.
Computing overlaps, our results are remarkable --- in a Hilbert space of dimension $10^8$, so $\dim\mathcal{H}\sim5\cdot10^6$ per momentum, we find a $99\%$ overlap in the flatband limit and an overlap $\gtrsim91\%$ throughout the entire FCI-SF transition for the $\mathbf{Q}=(0,0)$ VMC ansatz. The function $g_{\alpha\beta}(\mathbf{x},\mathbf{y})$ over which we optimize contains $d=567$ independent parameters. This gives $\frac{\dim\mathcal{H}}{d}\sim 10^4$ and the high overlaps indicate that the WF ansatz form captures the physics well. We show energies for both the best paired VMC state at $\mathbf{Q}=(0,0)$ and at $\mathbf{Q}=(\pi,0)$. While the latter is equally good in the FCI limit, it is high in energy in the SF phase because of its momentum. We also show the energies of the determinant-product state --- that is assuming independent partons but still using the full VMC technology. While the energy is good in the flatband limit and the wavefunction only has $d'=54$ parameters, Fig.~\ref{fig:checkerboard_ed_comparison} shows that this ansatz fails to captures the energetics of the SF state, demonstrating the need for the paired ansatz. In Sect.~\ref{sup:DMRG} of the Supplement~\cite{Supplement} we also compare our method to the DMRG results of Ref.~\cite{04_meng_numerics} on the honeycomb lattice,  with similar conclusions. While overlaps are considered to be a gold-standard for comparison of wavefunctions to numerics, finite size effects should be treated very carefully.  We have seen no indication of behavioral change at larger systems sizes and in Sect.~\ref{sup:mf_bcs_optim} of the Supplement \cite{Supplement}, we show how similar wavefunctions can be generated for arbitrary systems sizes without indication of any new gap closings.

We find numerically that only terms of the form $\MFEVs{f_1^\dagger f_1}$,$\MFEVs{f_2^\dagger f_2}$ and $\MFEVs{f_1f_2}$ obtain a significant nonzero expectation value. So while the ansatz Eq.~\ref{eq:general_MF_ham} is able to explore a variety of mean-field states, including ones with the gauge group broken to $\mathbb{Z}_2$, our results suggest that states with a $U(1)$ gauge group are favoured in this model. Our FCI state is topologically identical to that of Ref.~\cite{05_barkMcGreevyTheory}. 
As discussed in Sect.~\ref{sup:sf_field_theroy} of \cite{Supplement}, the nonzero $\MFEVs{f_1f_2}$ keeps us in the same SF phase, but improves energetics and allows for a larger SF weight. In fact in Sect.~\ref{sup:mf_bcs_optim} of the Supplement \cite{Supplement}, we show that most of the results in Fig.~\ref{fig:checkerboard_ed_comparison} can be recovered by an explicitly $U(1)$-invariant short-range parton MF Hamiltonian analogous to Eq.~\ref{eq:general_MF_ham}.

With a periodic structure in $g_{\alpha\beta}(\mathbf{x},\mathbf{y})$, we may define a MF BdG parton Chern number $\chern^\text{BdG}$ as discussed in the Supplement~\cite{Supplement}, Sect.~\ref{sup:bdgchern}. In the limit of decoupled partons, we have $\chern^\text{BdG}=2(\chern^\text{P}_1+\chern^\text{P}_2)$, suggesting $\chern^\text{BdG}=4$ for FCI and $\chern^\text{BdG}=0$ for SF. A jump of $\chern^\text{BdG}$ by four units is natural --- by the BdG nature, a gap closure at $\mathbf{k}$ implies one at $-\mathbf{k}$ and the projective translation argument then necessitates $\pm\mathbf{k}+(0,\pi)$, generically giving four closures as in Ref.~\cite{17_PhysRevLett.115.026802}. We observe such a jump at the transition, directly confirming the theoretical proposal of Refs.~\cite{05_barkMcGreevyTheory,15_Senthil_theory} with the transition described by a multiple parton gap closure. But our numerics show that to accurately capture the physics, a nonzero anomalous expectation $\Delta=\MFEV{f_1f_2}$ must be allowed.
While $\Delta$ is expected to be zero at the transition in the thermodynamic limit, we argue in Sect.~\ref{sup:sf_field_theroy} of \cite{Supplement} that its fluctuations affect details of the transition, including a re-normalization of the transition point, which we also observe numerically in an extrapolation to large system sizes.


In  Supplement\cite{Supplement}, Sect.~\ref{sup:SCMF}, we present a self-consistent MF method, informed by our above results which allows us to work with systems large enough to clearly see the quadruple band closure.  There are scenarios (explained in detail in Supplement~\cite{Supplement}, Sect.~\ref{sup:bdgchern}) where $\chern^\text{BdG}$ can 
change in steps of 2 \textit{without} breaking translational symmetry and thus we can have a phase with $\chern^\text{BdG}=2$ which is different from what has been proposed previously, as it can only emerge from a paired-parton ansatz. However, our numerics have not found examples where this is energetically favorable.

\begin{figure}
    \centering
    \includegraphics[width=3.5in]{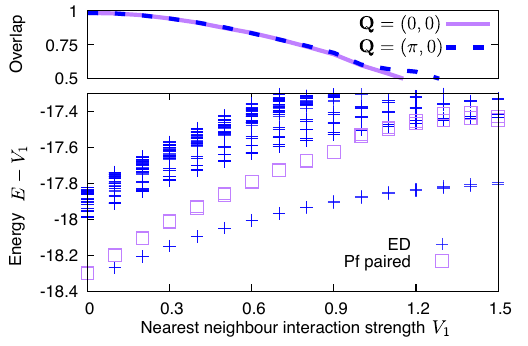} 
    \caption{Energy and overlap comparison of our paired Pf VMC ansatz to exact diagonalization for a finite nearest neighbor repulsion $V_1$ in the FCI phase at Bandwidth 0.09, showing good overlap for both ground states for weak interaction. The overlap becomes worse as $V_1$ approaches unity, the scale of the hopping parameters and band gap.
    }
    \label{fig:checkerboard_ed_comparison_int}
\end{figure}

We have so far considered models with the on-site hard-core constraint being the only interaction. We now add a repulsive density-density interaction $\hat{V}=\frac12V_1\sum_{\langle \mathbf{x},\mathbf{y}\rangle}\hat{n}_\text{B}(\mathbf{x})\hat{n}_\text{B}(\mathbf{y})$ between neighboring sites $\langle \mathbf{x},\mathbf{y}\rangle$. We show the energy and WF overlap agreement with exact results in Fig.~\ref{fig:checkerboard_ed_comparison_int} as a function of the interaction strength $V_1$.
While the agreement is good for relatively weak $V_1\lesssim \text{max}(|t|)=1$, at $V_1\sim1$ our trial states have an energy above the gap and a relatively poor ground-state overlap. Throughout the variation, the two topological ground states remain degenerate and our VMC procedure generates both equally well. Finally, we note that for large $V_1$, this system approaches a supersolid phase \cite{39_PhysRevLett.107.146803,40_PhysRevB.102.155120,lu2024vestigialgaplessbosondensity}, which may be part of the reason why our ansatz works relatively poorly there.

\textit{Conclusions.---} While it is very satisfying to find that the parton picture, once pairing is allowed, is so successful in describing both the SF and FCI phases as well as the transition between them, it is hard to avoid the question of generalizing this theory to other filling fractions. This could then lead to interesting studies of transitions between the $\nu=\frac13$ fermionic FCI and various CDW phases \cite{15_Senthil_theory}, or between $\nu=\frac13$ FCI and $\nu=\frac23$ fermionic FCI, a transition believed to be very similar to the one presented here, but with a tripled parton gap closure and with an emergent $SU(3)$ symmetry \cite{53_zaletel_vishwanath_PhysRevX.8.031015}. Unfortunately, the overlaps of analogously constructed fermionic FCI wavefunctions with the exact ground states are not as spectacular.
Finding physically motivated generalizations of Eq.~\ref{eq:mf_Pfaffian_form},\ref{eq:projected_bose_WF} having better overlap with the ED results would be an interesting future direction.

Note added --- Since our initial posting of this work, Ref.~\cite{wang2025emergentqed3bosoniclaughlin} has appeared online studying the same transition. Our results are consistent where they overlap.

\textit{Acknowledgments---} We thank the authors of Ref.~\cite{04_meng_numerics} for sharing some of their DMRG data (replotted in Supplement\cite{Supplement}). Exact diagonalization computations were performed using the DiagHam library. S.H.S. acknowledges support from EPSRC grant EP/X030881/1.

\bibliography{bibliography}

\onecolumngrid

\newpage
\clearpage

\begin{center}
    {\large\bf Supplementary Material for:}

\vspace*{10pt}

{\large \bf Paired Parton Trial States for the Superfluid-Fractional Chern Insulator Transition}

\begin{center}
   Tevž Lotrič and Steven H. Simon 
\end{center}

\vspace*{10pt}
    
\end{center}

\setcounter{section}{0}
\setcounter{equation}{0}
\setcounter{figure}{0}
\setcounter{table}{0}
\setcounter{page}{1}
\makeatletter
\renewcommand{\theequation}{S\arabic{equation}}
\renewcommand{\thefigure}{S\arabic{figure}}
\renewcommand{\thetable}{S\arabic{table}}

\newcommand{\onot}[1]{\mathcal{O}\left( #1 \right)}
\newcommand{\gk}{g_\mathbf{k}}
\newcommand{\Uk}{U_\mathbf{k}}
\newcommand{\Vk}{V_\mathbf{k}}
\newcommand{\csterm}[2]{\epsilon^{\mu\nu\lambda}#1_\mu\partial_\nu #2_\lambda}

\section{Lattice model hopping parameters \label{sup:hoppings}}
\begin{figure}[b]
    \centering
    \includegraphics[width=0.5\linewidth]{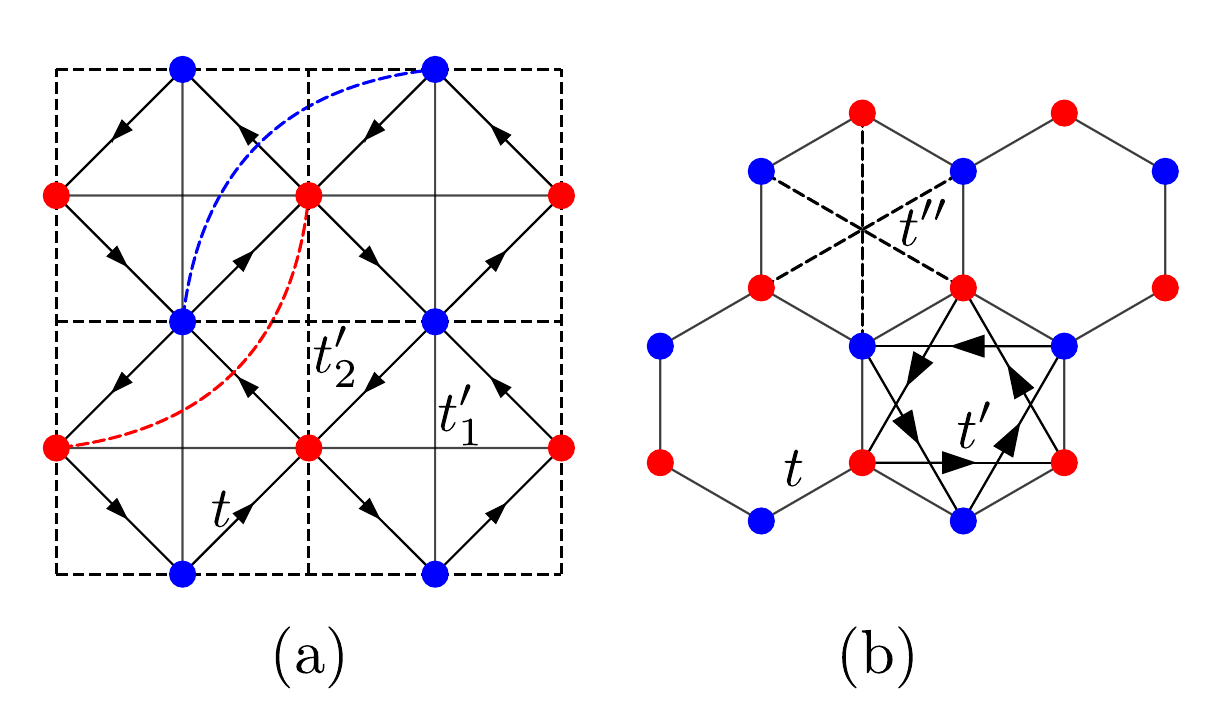}
    \caption{Hopping amplitudes for the checkerboard (a) and hexagonal (b) models. In (a), the curved lines appear between all diagonally separated pairs of like-colored points but are not shown for clarity. Similarly, in (b) both the $t'$ and $t''$ amplitudes appear in each plaquette.}
    \label{fig:hopping_sketch}
\end{figure}

The goal of this section is to clarify the exact hopping parameters $t(\mathbf{x},\mathbf{y})$ used for our FCI models. The checkerboard and hexagonal model are both illustrated in Fig.~\ref{fig:hopping_sketch}. Consider first (a), so the checkerboard model. Nearest-neighbors have a complex hopping $t$ (sign of the phase dictated by the arrow direction), while next-nearest neighbors are connected either by a solid or a dashed line for real hopping amplitude $t_1'$ or $t_2'$, respectively. Finally, there are next-next-nearest neighbor hoppings $t''$, indicated by the curved lines. For the particular parameter choice $t=e^{i\frac\pi4},t'_1=-t_2'=\frac{1}{2+\sqrt{2}}$ and $t''=\frac{1}{2+2\sqrt{2}}$, the lower band of this model is a very flat Chern band \cite{42_PhysRevLett.106.236803}. We call this the ``flatband limit'' but again note that the band is not perfectly flat.
To generate Fig.~\ref{fig:checkerboard_ed_comparison} in the main text, we tune away from this flatband limit into a band with a minimum at the $\Gamma$-point $\mathbf{k}=(0,0)$ in the Brillouin zone. This is done by the parameter choice $t=e^{i\frac{(\pi-\alpha)}4},t_1'=-t_2'=\frac{1}{2+\sqrt{2}}-\alpha,t''=\frac{1}{2+2\sqrt{2}}$. At $\alpha=0$, this is the flatband limit, while for $\alpha>0$ this produces a unique minimum at the desired momentum. This is just one of many possible tunings of the parameters that give this minimum, and we expect to see similar results for other variations as well. The interaction $V_1$ we discuss is added between all pairs of nearest neighbor sites, that is any pair of sites connected by a hopping amplitude $t$ in Fig.~\ref{fig:hopping_sketch}(a).

In Sect.~\ref{sup:DMRG} below, we also compare our results to the DMRG of Ref.~\cite{04_meng_numerics} on the hexagonal model, Fig.~\ref{fig:hex_vmc_dmrg}(b). Here $t$,$t'$ and $t''$ are the nearest, next-nearest and next-next-nearest neighbor hoppings, respectively. Only $t'$ is taken to be complex. Again it is known that when $t=1$, $t'=0.6e^{0.4\pi i}$ and $t''=-0.58$, the lower Chern band is very close to (but not exactly) flat. When $t''>-0.58$, a minimum at the $\Gamma$-point emerges, similar to the checkerboard case above \cite{04_meng_numerics}.

{
\section{Mean-field parton Hamiltonians for the paired state} \label{sup:mf_bcs_optim}
The only significant parton bilinear expectations in the paired state for our models are $\MFEVs{f_1^\dagger f_1}$, $\MFEVs{f_2^\dagger f_2}$  and $\MFEVs{f_1f_2}$, which all preserve an internal $U(1)$ gauge invariance. In this section, we restrict the other bilinears to zero. 
Further, we note that while both the determinant product and paired ansatzes discussed are free-fermion states for the partons, it is not immediately clear whether they can emerge from a local Hamiltonian.

To test both of these issues, we consider paired states with the pairing function $g_{\alpha\beta}(\mathbf{x},\mathbf{y})$ derived from a local Hamiltonian. In particular, we consider the class of Hamiltonians of the form 
\begin{equation} \label{eq:constrained_hmf}
        H^\text{MF} =\sum_{\mathbf{x},\mathbf{y}}\left[t^\text{MF}_{11}(\mathbf{x},\mathbf{y}) f_1^\dagger(\mathbf{x})f_1(\mathbf{y})+t^\text{MF}_{22}(\mathbf{x},\mathbf{y}) f_2^\dagger(\mathbf{x})f_2(\mathbf{y}) +\Delta^\text{MF}_{12}(\mathbf{x},\mathbf{y}) f_1(\mathbf{x})f_2(\mathbf{y}) + \text{h.c.}\right]
\end{equation}
which clearly perserves the $U(1)$ gauge invaraince $f_1\rightarrow e^{i\alpha}f_1$ and $f_2\rightarrow e^{-i\alpha}f_2$. We further enforce that the mean-field hoppings may be written as $t^\text{MF}_{11}(\mathbf{x},\mathbf{y})=t^0_{11}(\mathbf{x},\mathbf{y}) e^{ia(\mathbf{x},\mathbf{y})}$ where $t^0_{11}$ respects the full bosonic translational symmetry, and $a(\mathbf{x},\mathbf{y})$ are link variables enforcing $\pi$-flux per unit cell, with an analogous relation for $t_{22}^\text{MF}$ and $\Delta_{12}^\text{MF}$. Note that this is a stricter translation symmetry requirement than only translational invariance in a doubled unit-cell.

With these constraints, Eq.~\ref{eq:constrained_hmf} may be readily diagonalized, giving us a pairing function $g$ (details in Sect.~\ref{sup:bdgchern}) which we may then use to sample the bosonic wavefunction as discussed above. We may then vary the effective mean-field Hamiltonians to optimize the true bosonic energy. We compare the results to the unconstrained Pf wavefunction and to ED in Fig.~\ref{fig:hbscmf}. Surprisingly, we find that most of the results can be reproduced by an extremely short-range $H^\text{MF}$, allowing for nonzero $\Delta^\text{MF}_{12}$ only on-site and for nonzero $t^\text{MF}_{11},t^\text{MF}_{22}$ only up to next-nearest neighbour hopping. So the effective mean-field Hamiltonian can have a shorter range than the underlying true bosonic Hamiltonian. 

\begin{figure}
    \centering
    \includegraphics[width=0.65\linewidth]{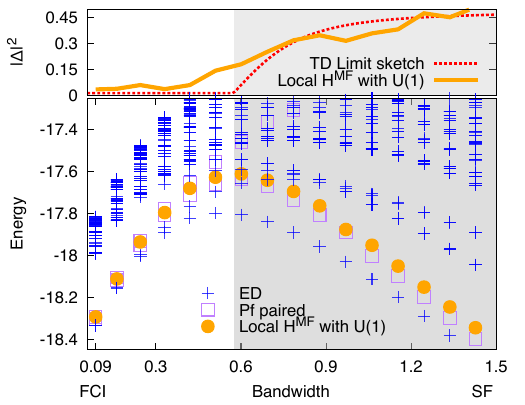} 
    \caption{  Comparison of the energies obtained from a short-range $U(1)$ invariant effective mean-field Hamiltonian (Eq.~\ref{eq:constrained_hmf}) compared to the (strictly more general) unconstrained Pfaffian ansatz and exact diagonalization. The local $U(1)$ mean-field Hamiltonian correctly reproduces most of the energy variation, despite only short-range hopping and on-site pairing being allowed. The top shows the strength of the pairing $|\Delta_{12}^\text{MF}|^2$ in the effective parton mean-field Hamiltonian. The scale is chosen such that $t_{11,22}^\text{MF}$ have average magnitude unity -- this indicates that $|\Delta|\sim0.7|t|$ in the SF. The transition is not sharply defined, which we believe to be a finite system-size effect. The red dotted line indicates the expected generic behavior in the large-system thermodyanamic (TD) limit. The behaviour of our ansatz in the TD limit is analysed more carefully in Fig.~\ref{fig:tdlimscaling}.
    }
    \label{fig:hbscmf}
\end{figure}

By explicitly enforcing the $U(1)$ gauge invarance and locality of the parton mean-field Hamiltonian, we still find excellent agreement with the exact ground states, suggesting that our $U(1)$ paired parton ansatz is truly the appropriate one. This approach has several other upsides -- the number of independent complex variational parameters is now reduced to $d''=20$, much fewer than the $d=567$ and $d'=54$ used for the generic paired and determinant product ansatzes above. Furthermore, this ansatz allows one to find the effective $H^\text{MF}$ on a smaller system and then trivially increase the system size to find good wavefunctions for larger sizes. This allows us to test larger systems sizes and look for new gap closings or other changes of behavior that might occur (in fact we find that our small system size numerics are very representative of the behavior at larger system size).

\section{Effective field theory of the superfluid state} 
\label{sup:sf_field_theroy}

In this section, we present an effective field-theoretical description of the superfluid phase described by our paired ansatz. While the generic pairing ansatz (Eq.~\ref{eq:general_MF_ham} of main text) allows for any fermionic bilinear to obtain a nonzero expectation, given our numerical results it is sensible to focus on the cases where only $t^\text{MF}_{11},t^\text{MF}_{11}$ and $\Delta_{12}^\text{MF}$ are nonzero, so an ansatz of the form Eq.~\ref{eq:constrained_hmf}. So this section focuses on cases where $\MFEVs{f_1^\dagger f_1}$, $\MFEVs{f_2^\dagger f_2}$ and $\MFEVs{f_1f_2}$ are allowed to be nonzero. This is a necessary discussion, as in our descriptions, there are two ``pairing mechanisms'' for the partons. One is simply the mean-field pairing, which can result by inserting terms such as $\Delta^\dagger f_1f_2 \subset H^\text{MF}$, with $\Delta$ playing the role of an order parameter. The other mechanism, described in \cite{05_barkMcGreevyTheory} results from fluctuations of the internal $U(1)$ gauge field $a_\mu$ and is present even when there is no mean-field pairing present. The order parameter of this mechanism is related to the instanton $\mathcal{M}_a$ inserting $2\pi$ flux of $a_\mu$. To understand the phase of our paired ansatz, we must address the question of how these two order parameters relate to each other. In Sect.~\ref{sub:bm_review}, we review how the gauge-field fluctuations effectively pair the partons into a SF state. Then in Sect.~\ref{sub:paired_ft}, we argue that the true ground state in this phase generically has both $\langle \mathcal{M}_a\rangle\neq0$ and $\langle \Delta\rangle\neq0$, but the two order parameters together produce only one single superfluid mode. Such a state may be continiously deformed to one with $\Delta=0$, meaning it is in the same unversality class as the superfluid described by \cite{05_barkMcGreevyTheory}. But it is generally a better description of that phase, allowing for a larger superfluid weight.

\subsection{Review of pairing by gauge-field fluctuations} \label{sub:bm_review}
The discussion by Barkeshli and McGreevy \cite{05_barkMcGreevyTheory} starts by assuming that the partons $f_{1,2}$ are independent at the mean-field level and independently fill bands with Chern number $\chern_1^\text{P}=1$ and $\chern_2^\text{P}=-1$. They are coupled to gauge fields $\frac12A_\mu+a_\mu$ and $\frac12A_\mu-a_\mu$ respectively, where $A_\mu$ is the external EM field and $a_\mu$ is the internal gauge field which has been assumed to be Higgsed $SU(2)$ down to $U(1)$. We describe this phase by
\begin{equation}
    \mathcal{L} = f_1^\dagger \mathcal{D}_{\frac12A+a}^{(1)}f_1 + f_2^\dagger\mathcal{D}_{\frac12A-a}^{(-1)}f_2 \label{eq:fermi_eff_l_no_pair}
\end{equation}
where $\mathcal{D}^{(\chern)}_b$ is a kinetic term putting the fermion into a Chern-$\chern$ band and minimally coupling it to a gauge field $b_\mu$. Integrating out the fermions gives
\begin{equation} \label{eq:bm_sf_lagrangian}
    \mathcal{L}=\frac{1}{4\pi}\left(\frac12A+a\right)d\left(\frac12A+a\right)-\frac{1}{4\pi}\left(\frac12A-a\right)d\left(\frac12A-a\right)+\ldots = \frac{1}{2\pi}Ada+\ldots,
\end{equation}
where $adb=\varepsilon^{\mu\nu\rho}a_\mu\partial_\nu b_\rho$ denotes the Chern-Simons term and $\ldots$ indicate less RG-relevant terms. Due to the Chern-Simons term, the instanton $\mathcal{M}_a$ carries unit charge under $A_\mu$, and it has been argued that the low-energy theory can include $\mathcal{M}_a f_2^\dagger f_1^\dagger$, giving the interpretation of $\mathcal{M}_a$ as the order parameter of this superfluid \cite{05_barkMcGreevyTheory}. In the following, it will be convenient to re-write Eq.~\ref{eq:bm_sf_lagrangian} using the $U(1)$ boson-vortex duality \cite{PESKIN1978122,Halperin_1981_u1_duality,Seiberg_2016}. Taking Eq.~\ref{eq:bm_sf_lagrangian} as the vortex side, we may re-write it as (in Euclidian time)
\begin{equation} \label{eq:bm_sf_after_duality}
    \mathcal{L}=|D_A\phi|^2 - r_\phi|\phi|^2+|\phi|^4.
\end{equation}
for a complex bosonic field $\phi$. Here, $D_A=\partial_\mu-iA_\mu$ and we have $r_\phi>0$, so $\phi$ is in the condensed phase. Thus Eq.~\ref{eq:bm_sf_after_duality} and by duality also Eq.~\ref{eq:bm_sf_lagrangian} represent a superfluid. In the duality, the instanton $\mathcal{M}_a$ is mapped onto the operator $\phi$, supporting the above identification of the instanton as an order parameter. Here and in what follows, we do not keep track of the pre-factors of quartic terms, noting that they do not matter at our level of analysis.

\subsection{Extension to paired parton states} \label{sub:paired_ft}
Going back to the original bosonic Hamiltonian in Eq.~\ref{eq:bose_ham}, $  H^\text{B}=\sum t({\mathbf{x},\mathbf{y}})b^\dagger(\mathbf{x})b(\mathbf{y})$ and inserting $b=f_1f_2$, we get $H=\sum t({\mathbf{x},\mathbf{y}})f_2^\dagger(\mathbf{x})f_1^\dagger(\mathbf{x})f_1(\mathbf{y})f_2(\mathbf{y})$. Say that we were to decouple \textit{some} of the terms in the pairing channel by introducing a Hubbard–Stratonovich field $\Delta$. This would then give us $H=H_\Delta+\sum \tilde{t}({\mathbf{x},\mathbf{y}})f_2^\dagger(\mathbf{x})f_1^\dagger(\mathbf{x})f_1(\mathbf{y})f_2(\mathbf{y}) + \sum \Delta^\dagger(\mathbf{x})f_1(\mathbf{x})f_2(\mathbf{x})+\text{h.c.}$, where $\Delta$ is a dynamical field and in general $\tilde{t}(\mathbf{x},\mathbf{y})\neq t(\mathbf{x},\mathbf{y})$. The term $H_\Delta$ depends only on $\Delta$ and is the result of the Hubbard-Stratonovich decoupling. Note that $\Delta$ is to be determined self-consistently, so doing this has not yet enforced any mean-field pairing, but has merely allowed for its possibility. Furthermore, $\Delta$ is not charged under the internal $U(1)$ gauge field $a_\mu$ so even in the case that $\Delta$ condenses the $U(1)$ gauge field $a_\mu$ survives. Continuing the analysis within the parton mean-field framework as before and assuming that the new hoppings $\tilde{t}$ still lead to a mean-field phase with $(\chern_1^\text{P},\chern_2^\text{P})=(1,-1)$, we get the effective description
\begin{align}
    \mathcal{L} &= f_1^\dagger \mathcal{D}_{\frac12A+a}^{(1)}f_1 + f_2^\dagger\mathcal{D}_{\frac12A-a}^{(-1)}f_2 + \Delta f_2^\dagger f_1^\dagger + \Delta^\dagger f_1f_2 +\mathcal{L}_\Delta \label{eq:fermi_eff_l_with_pairing}\\
    \mathcal{L}_\Delta&= |D_A\Delta|^2 + r_\Delta|\Delta|^2 +|\Delta|^4+\ldots \label{eq:eff_l_delta}
\end{align}
The form of $\mathcal{L}_\Delta$ is inferred based on symmetries and the fact that $\Delta$ carries unit charge under $A_\mu$, and no charge under $a_\mu$. This effective theory approach however cannot give us information about the sign of $r_\Delta$. The next step is to integrate out the fermions as we did in Eq.~\ref{eq:fermi_eff_l_no_pair}. The crucial subtlety is that, as discussed above, the operator $\mathcal{M}_a f_2^\dagger f_1^\dagger$ must be included in the effective low-energy theory. So both $\Delta$ and $\mathcal{M}_a$ pair to the same fermionic bilinear, $f_2^\dagger f_1^\dagger$. It is thus natural to expect terms $\sim \Delta^\dagger\mathcal{M}_a$ to appear in the effective theory. Working in a small-$\Delta$ expansion, we infer an effective theory 
\begin{equation} \label{eq:delta_and_inst}
    \mathcal{L}= \frac{1}{2\pi}Ada+c\Delta^\dagger\mathcal{M}_a+c^*\Delta\mathcal{M}_a^\dagger+ |D_A\Delta|^2 + r_\Delta|\Delta|^2 + \mathcal{L}^{(4+)}(\mathcal{M}_a,\Delta)
\end{equation}
We may alternatively justify the terms $c\Delta^\dagger\mathcal{M}_a$ by simply noting that they are allowed (Sect. 3.2 of \cite{Seiberg_2016}), meaning that we expect them to appear generically. The term $\mathcal{L}^{(4+)}$ contains all terms of quartic and higher order in $\mathcal{M}_a$ and $\Delta$, which we assume are such that Eq.~\ref{eq:delta_and_inst} is bounded (e.g. $|\Delta|^4\subset\mathcal{L}^{(4+}$, \ldots). Proceeding with a particle-vortex duality on $da$ as above, we find the action
\begin{equation} \label{eq:paired_sf_after_pv_Dual}
    \mathcal{L}=|D_A\phi|^2 + |D_A\Delta|^2 -r_\phi|\phi|^2+r_\Delta|\Delta|^2+c^*\phi^\dagger\Delta+c\Delta^\dagger\phi  + \tilde{\mathcal{L}}^{(4+)}(\phi,\Delta)
\end{equation}
where we used that the duality maps $\mathcal{M}_a\leftrightarrow\phi$ \cite{Seiberg_2016} and now $\tilde{\mathcal{L}}^{(4+)}\sim\mathcal{L}^{(4+)}+|\phi|^4$. Because $r_\phi>0$, the point $\phi=0,\Delta=0$ cannot be the minimum of the potential. It can be shown that for general $c\neq0$ the minimum is traced out by $\phi=\alpha e^{i\varphi},\Delta=\beta e^{i\varphi+i\delta}$ when $\varphi$ is varied for some fixed $\alpha,\beta,\delta\in\mathbb{R}$. Long-wavelength fluctuations of $\varphi$ describe the usual superfluid mode. When $c\neq0$, all other fluctuations are gapped, namely the fluctuations of the magnitudes $\alpha,\beta$ and the relative phase fluctuations $\delta$. The low-energy effective theory is then simply
\begin{equation} \label{eq:final_double_sf}
    \mathcal{L}=(|\alpha^2|+|\beta|^2)\left(\partial\varphi-A\right)^2,
\end{equation}
evidently describing a charged superfluid. Regardless of the sign of $r_\Delta$, the minimum has this form (as long as $c\neq0$) and it can be shown that the minimum has $\langle |\Delta|\rangle\neq0$ even when $r_\Delta>0$. Furthermore, the pre-factor in Eq.~\ref{eq:final_double_sf} suggests that both $\phi$ and $\Delta$ contribute to the superfluid weight -- we may interpret this paired construction as a way of achieving a larger superfluid weight than what is possible by only the instanton mechanism. 

So while our construction may look like it has two independent order parameters, $\mathcal{M}_a\leftrightarrow\phi$ and $\Delta$, they are coupled by the term $c\Delta^\dagger\phi$, which makes it so that both together only have one superfluid mode. The crucial ingredient for all this to work is $c\neq0$, but as argued above, we believe this to be true away from fine-tuned points. 

It should be noted that any point in this phase may be connected to $\langle|\Delta|\rangle=0$ without encountering a phase transition -- we can keep $r_\phi>0$, then first move to a point $r_\Delta>0$, and then take $c\rightarrow0$. While such a point is fine-tuned, it shows the equivalence between the resulting SF phase in the paired and unpaired parton constructions.

\subsection{Implications for the transition}

The above has established a field-theoretical framework for understanding the superfluid phase in the paired parton construction. The aim of this section is to complement the finite system-size microscopic numerical results we found for the transition with a field-theoretical picture in the thermodynamic limit. Fig.~\ref{fig:hbscmf} shows how $|\langle\Delta\rangle|^2$ evolves as a function of the bandwidth. We should expect that in the thermodynamic limit, $|\langle \Delta\rangle|=0$ everywhere in the FCI phase in order to not break the external $U(1)$ symmetry and that $|\Delta|$ only starts to grow in the SF phase, where the charge $U(1)$ is already broken by the instanton pairing mechanism. This means that in the thermodynamic limit the transition point still has $\Delta=0$, which means a more detailed discussion of how pairing influences the transition is needed.

The above reasoning suggests that the transition manifests at the mean-field level as both a parton Chern number change \textit{and} a spontaneous symmetry-breaking transition for $\Delta$. This implies that $\Delta$ becomes soft at the transition, meaning that we should expect significant fluctuations in $\Delta$ at and near the transition, despite it still being zero on average at the transition point. Furthermore, the FCI-SF transition may be characterized by critical exponents on the SF side. There, the mean-field pairing $\Delta$ is non-zero, meaning that the parton pairing could have implications on the nature of the state observed there and consequently on critical exponents obtained.


To see this more explicitly, consider the SF side of the theory, Eq.~\ref{eq:paired_sf_after_pv_Dual}, as we approach the transition. The quadratic potential terms may be written as $(\phi^\dagger,\Delta^\dagger)M\begin{pmatrix}
    \phi\\ \Delta
\end{pmatrix}$ with $M=\begin{pmatrix}
    -r_\phi & c^*\\ c & r_\Delta 
\end{pmatrix}$. In the SF phase, $M$ has a negative eigenvalue $\lambda_-<0$, implying that some combination of $\phi,\Delta$ condenses, as discussed above. But as we approach the transition with $\lambda_-\rightarrow0^-$, the fluctuations in the associated eigenvector $Mv_-=\lambda_-v_-$ become soft. In general ($c\neq0$), $v_-$ has components in both $\phi,\Delta$, showing that one should expect critical fluctuations in both quantities as the transition is approached from the SF side. From the continuity of the transition, we should expect to see significant fluctuations of $\Delta$ also when approaching the transition from the FCI side.

\begin{figure}
    \centering
    \includegraphics[width=0.6\linewidth]{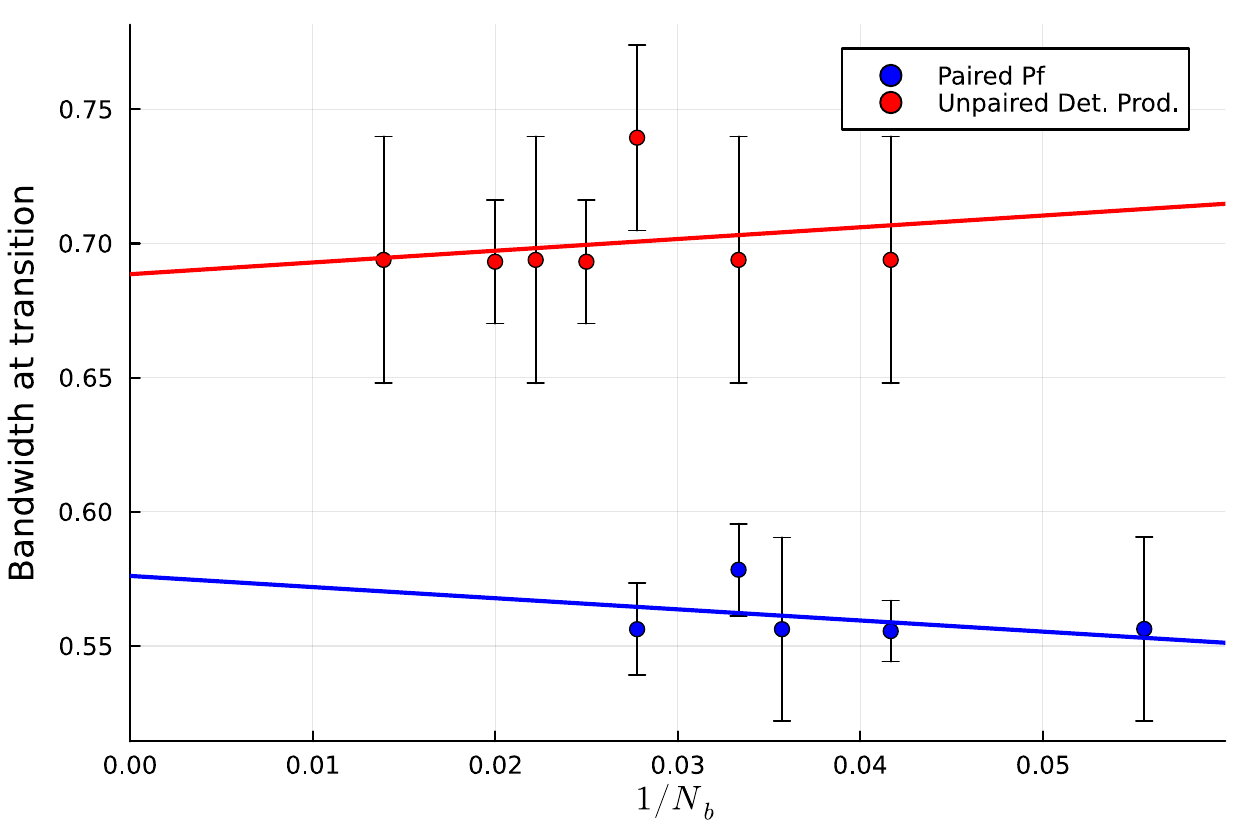}
    \caption{The dependence of the bosonic band width of the FCI-SF transition in the checkerboard model (determined by the change in parton Chern numbers) on the number of bosons $N_b$ in the system for the paired and unpaired ansatz. Extrapolation to the thermodynamic limit gives $w_\text{unpaired}=0.69\pm0.05$ and $w_\text{paired}=0.58\pm0.04$. This analysis implies a difference $\delta w=w_\text{unpaired}-w_\text{paired}=0.11\pm0.06$ for large systems, demonstrating that mean-field parton pairing has a meaningful influence on the nature of the transition even in the thermodynamic limit.} 
    \label{fig:tdlimscaling}
\end{figure}

Furthermore, accounting for the mean-field pairing $\Delta$ re-normalizes the transition point: in the following, we look only at  a mean-field picture and assume $r_\Delta>0$. If we decouple the mean-field pairing (send $c\rightarrow0$), the state is in the SF phase as long as $r_\phi>0$. But when $c\neq0$, we are in the SF if $r_\phi\geq -|c|^2/r_\Delta$, a less strict condition. It is easier for a combination of $\phi,\Delta$ to remain condensed than just $\phi$ itself. So allowing for a nonzero $\Delta$ changes the transition point even in the thermodynamic limit, implying that accounting for it is crucial to correctly capture the transition. This can be confirmed numerically -- in Fig.~\ref{fig:tdlimscaling}, we compare the bandwidth at which the SF-FCI transition is observed for the two ansatzes at various system sizes. Extrapolation to the thermodynamic limit suggests that the transition point is indeed different between the two ansatzes, demonstrating the need for pairing. We should caution that Eq.~\ref{eq:paired_sf_after_pv_Dual} applies only in the SF phase -- while we may use that to comment on the location of the transition, Eq.~\ref{eq:paired_sf_after_pv_Dual} does not apply at the transition itself. To understand that further, one would have to find an appropriate paired-parton generalization of the QED$_3$ theory of \cite{05_barkMcGreevyTheory}.

This all suggests that including $\Delta$ is important to understand the microscipic energetics and to correctly capture the wavefunction at and around the transition.  Our analysis is limited to the mean-field level. We should expect that if one were able to do the integral over the Chern-Simons field $a_\mu$ exactly, the result should be exact regardless of whether $\Delta$ is included. But the idea behind the parton construction is to obtain a non-trivial mean field around which to expand. The paired parton ansatz significantly improves the quality of the mean field approximation, but the exact nature of the transition would only become apparent after the fluctuations are accounted for, for example following the $1/N_f$ perturbative expansion in \cite{05_barkMcGreevyTheory}. Including $\Delta$ could help capture the fluctuations more accurately in such expansions, but the question of how relevant such fluctuations are at the critical point and by how much they alter the resulting critical exponents remains open. Aspects of this transition have been recently analyzed numerically in \cite{wang2025emergentqed3bosoniclaughlin}.

To summarize, we developed an effective field theory framework to describe the paired parton states which contain the particular pairing term $\Delta f_2^\dagger f_1^\dagger$ and which do not break $U(1)$ internal gauge invariance. We found that within the superfluid phase, the saddle point of this theory generally involves some nonzero mean-field pairing parameter $\Delta$, meaning that to describe the superfluid, we should always allow for some mean-field pairing. Our analysis shows that the mean-field pairing and gauge-fluctuation-mediated mechanisms both work together to produce a single superfluid. {Furthermore, while $|\langle \Delta \rangle|=0$ in the FCI phase and at the transition, the parton pairing $\Delta$ is still important at the transition -- allowing for it re-normalizes the transition point, and we should expect fluctuations in $\Delta$ to become soft at the transition.

\section{Algorithm details and complexity \label{sup:alg_trickery}}
Our VMC algorithm follows that of \cite{35_PhysRevB.108.245128,33_doi:10.1126/science.aag2302} closely. In order to minimize $E[\phi]=\frac{\braketmatrix{\phi}{H^\text{B}}{\phi}}{\braket{\phi}{\phi}}$, we must be able to evaluate the local energy $E_L(\{\mathbf{r}\})=\phi^{-1}(\{\mathbf{r}\}) \hat{H}^\text{B} \phi(\{\mathbf{r}\})$ as well as the derivative of our wavefunction with respect to the variational parameters $\mathcal{O}_g=\phi^{-1}(\{\mathbf{r}\}) \frac{\partial}{\partial g} \phi(\{\mathbf{r}\})$. With that, $\langle E_L \mathcal{O}_g^*\rangle-\langle E_L\rangle\langle \mathcal{O}_g^*\rangle$ determines the derivative of the energy $E[\phi]$ with respect to $g$. With these quantities, we use the ADAM stochastic optimizer \cite{37_kingma2017adammethodstochasticoptimization} to update our parameters.

In this section, we wish to discuss the computational complexity of our algorithm at each sampling point in the VMC algorithm. Most of the complexity is in evaluation the wavefunction $\phi=\text{Pf}A$ which is $\onot{N^3}$ for $N$ particles. With the same $\onot{N^3}$ complexity, we also evaluate $A^{-1}$. Now we may use the fact that $\partial_g\phi=\partial_g\text{Pf}A=\frac12\text{tr}\left[A^{-1}\partial_g A\right]$ to evaluate the derivative with respect to any component of $g$ in $\mathcal{O}(N^2)$ at worst. For the $\onot{N}$ variational parameters, this leads to a $\onot{N^3}$ complexity for gradient computation at worst. In practice, the computation is $\onot{N^2}$ as for any variational parameter, many elements of $\partial_gA$ are zero.
The evaluation of the local energy $E_L$ is more problematic, taking up most of the computational effort -- the Hamiltonian Eq.~\ref{eq:bose_ham} contains $\onot{N}$ hopping terms. The inclusion of each hopping term means re-evaluation of $\phi$ at a position $\{\mathbf{r}'\}$ in which one boson has moved relative to $\{\mathbf{r}\}$. Naively, each re-evaluation of $\phi$ costs $\onot{N^3}$, leading to an $\onot{N^4}$ cost for the evaluation of $E_L$. But it turns out that knowing $A^{-1}$ at $\{\mathbf{r}\}$, we may use the Woodbury matrix identity as in Ref.~\cite{47_Xu_2022} to compute a finite-rank update (we need rank-2 as a moving a boson means moving two partons) on the Pfaffian in $\onot{N^2}$. Thus the evaluation of the hopping part of $E_L$ is $\onot{N^3}$. Finally, note that density-density interaction terms are diagonal in position space and are thus computationally cheap, $\onot{N^2}$ to include.

Putting all of this together, it becomes evident that each step has a computational cost of $\onot{N^3}$, with the evaluation of $A^{-1}$ and $E_L$ being bottlenecks. Since we make local moves in our sampling process, the new inverse matrix may be evaluated as an update of the previous one in $\onot{N^2}$, but the $\onot{N^3}$ evaluation of $E_L$ cannot be sped up further. Finally, we note that in the uncorrelated determinant-product state, the algorithm can be sped up to $\onot{N^2}$ per step, as all we need for $E_L$ are rank-1 determinant updates, which may be computed in $\onot{N}$ using Cramer's rule.

\section{Comparison to DMRG on the hexagonal model \label{sup:DMRG}}
In Fig.~\ref{fig:checkerboard_ed_comparison}, we compared our ansatz on one particular lattice model to exact diagonalization. To show it works elsewhere as well, we compare it to cylinder-DMRG done on the hexagonal flatband model by Ref.~\cite{04_meng_numerics}. We thank the authors for sharing this data.
While the DMRG was done on a $30\times8$ cylinder, we choose to do the VMC on an $8\times8$ torus, comparing the energy per boson. The results are in Fig.~\ref{fig:hex_vmc_dmrg}, where the next-next-nearest neighbor hopping amplitude $t''$ is varied to induce a FCI-SF transition.

\begin{figure}
    \centering
    \includegraphics[width=3.5in]{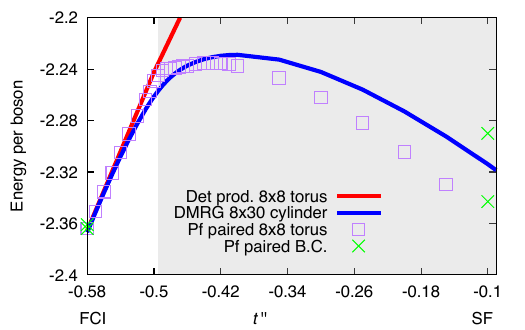}
    \caption{Comparison of our VMC method to DMRG data obtained in \cite{04_meng_numerics}. While a determinant product ansatz captures the FCI phase well, it fails near the transition and in the SF phase. The Pf paired ansatz works much better there. Due to the different boundary conditions, the Pf paired ansatz produces an energy lower than the DMRG, but within the estimated finite-size effect, the two are consistent.}
    \label{fig:hex_vmc_dmrg}
\end{figure}

The results show some consistency, but a direct comparison is difficult due to the different boundary conditions imposed: while the DMRG is naturally well-suited to work with open boundary conditions, our VMC would need substantially more variational parameters when translational invariance is given up in one of the directions. While not impossible, this would be computationally very expensive. We thus compare open-boundary cylinder DMRG to VMC on a torus. To estimate the effect of the boundary conditions, we also run VMC with twisted bosonic boundary conditions around each loop and use the observed variation of energy as an estimate of the boundary effect --- it is clearly much larger in the SF phase. This is entirely a finite-size effect, as the boundary conditions force the bosons to move away from the minimum in the dispersion and would go away in the thermodynamic limit, similar to the gap on the SF side in Fig.~\ref{fig:checkerboard_ed_comparison} in the main text. The two crucial findings of Fig.~\ref{fig:checkerboard_ed_comparison} of the main text are repeated here --- the Pf paired ansatz matches DMRG well in both phases, while the determinant product state only works well in the FCI limit.

The Pf paired variational ansatz finds the transition, here defined as the point where $\chern^\text{BdG}=4\rightarrow\chern^\text{Bdg}=0$ at a point which agrees with the result of the scaling analysis in \cite{04_meng_numerics}, $t''\sim-0.5$. The determinant product state also sees a transition (there $\chern^\text{P}_2=1\rightarrow\chern^\text{P}_2=-1$), but the transition is found at $t''=-0.4$. So not only are the energetics poor on the SF side, but the determinant-product also finds the incorrect transition point.

\section{Paired state details \label{sup:bdgchern}}

\subsection{Bogoliubov transformation and the mean-field wavefunction form}
Consider a model with $N_U$ sites per extended partonic unit cell (so eg the doubled unit cell for our problem with two sites per original unit cell has $N_U=4$) and with each physical particle split into $K$ partons ($K=2$ in our bosonic case). From the above discussion, we should really then treat the mean-field problem as having $D=KN_U$ parton sites per unit cell. Let $\rho,\sigma=1\ldots D$ be indices in these unit cells. Fourier-transforming, we get fermionic operators $f_{\sigma, \mathbf{k}}$ and Eq.~\ref{eq:general_MF_ham} of the main text may be written in BdG form as
\begin{equation} \label{eq:bdg_form}
    H_\text{MF}=\sum_\mathbf{k} \Psi^\dagger_\mathbf{k} \begin{pmatrix}
        T(\mathbf{k}) & \Delta(\mathbf{k}) \\
        \Delta^\dagger(\mathbf{k}) & -T^*(\mathbf{-k})
    \end{pmatrix}\Psi_\mathbf{k}
\end{equation}
where $\Psi_\mathbf{k}=(f_{1,\mathbf{k}},\ldots,f_{D,\mathbf{k}},f^\dagger_{1,\mathbf{-k}},\ldots,f^\dagger_{D,-\mathbf{k}})$ and $\Delta(-\mathbf{k})=-\Delta^T(\mathbf{k})$. Both $T({\bf k})$ and $\Delta({\bf k})$ are $D\times D$ matrices in this sublattice space. So the matrix in Eq.\ref{eq:bdg_form} is $2D\times 2D$ so we get $2D$ BdG bands. From the BdG form, we expect the MF GS $\ket{\Omega}$ to fill $D$ of these bands. What follows is a generalization of Read and Green's  logic \cite{46_Read_Green} to multiple lattice sites. Following Read and Green, the $D$ empty bands correspond to operators
\begin{equation} \label{eq:bogoliubov_transform}
    \alpha_{l\mathbf{k}}=(U_\mathbf{k})_{l\sigma} f_{\sigma,\mathbf{k}} - (V_\mathbf{k})_{l\sigma}f_{\sigma,\mathbf{-k}}^\dagger
\end{equation}
for $l=1...D$. The Latin indices represent bands and the Greek indices represent (parton) lattice sites in the model.
The ground state is then determined by the constraint that $\alpha_{l\mathbf{k}}\ket{\Omega}=0$ for all $\mathbf{k}$ and $l$. It may be written as (prime on the product indicates to only multiply over one of each $\mathbf{k},-\mathbf{k}$ pair)
\begin{align} \label{eq:bdg_gs_form}
\ket{\Omega}&=\text{const}\prod_\mathbf{k}'\exp\left[(g_\mathbf{k})_{\sigma\rho}f_{\sigma,\mathbf{k}}^\dagger f^\dagger_{\rho,\mathbf{{-k}}}\right]\ket{\emptyset} \\
g_\mathbf{-k}&=-g_\mathbf{k}^T. \label{eq:g_antisym}
\end{align}
Here $\ket{\emptyset}$ is the state annihilated by all the $f$ operators. To see why this is the case, note first that
\begin{equation} \label{eq:f_on_bdg_vacuum}
    \begin{split}
f_{\beta,\mathbf{k}}\ket{\Omega}&=\text{const}f_{\beta,\mathbf{k}}\prod_\mathbf{q}'\exp\left[(g_\mathbf{q})_{\sigma\rho}f_{\sigma,\mathbf{q}}^\dagger f^\dagger_{\rho,\mathbf{{-q}}}\right]\ket{\emptyset}=\\
        &=\text{const}\left\{ \prod_{\mathbf{q}\neq \pm \mathbf{k}}'\exp\left[(g_\mathbf{q})_{\sigma\rho}f_{\sigma,\mathbf{q}}^\dagger f^\dagger_{\rho,\mathbf{{-q}}}\right] \right\} f_{\beta,\mathbf{k}}  \exp\left[(g_\mathbf{k})_{\sigma\rho}f_{\sigma,\mathbf{k}}^\dagger f^\dagger_{\rho,\mathbf{{-k}}}\right]\ket{\emptyset} \\
    \end{split}
\end{equation}
Then, writing $G=(g_\mathbf{k})_{\sigma\rho}f_{\sigma,\mathbf{k}}^\dagger f^\dagger_{\rho,\mathbf{{-k}}}$, we can see that $[f_{\beta,\mathbf{k}},G]=(g_\mathbf{k})_{\beta\rho}f^\dagger_{\rho,\mathbf{-k}}$ from which it follows (with $e^{-A}Be^A=B+[B,A]+\frac{1}{2!}[[B,A],A]+\frac{1}{3!}[[[B,A],A],A]+\ldots$) that  $f_{\beta,\mathbf{k}}e^G=e^G\left(f_{\beta,\mathbf{k}}+(g_\mathbf{k})_{\beta\rho}f^\dagger_{\rho,\mathbf{-k}}\right)$.Using that in Eq.~\ref{eq:f_on_bdg_vacuum} along with $f_{\beta,\mathbf{k}}\ket{\emptyset}=0$ and $e^Gf^\dagger_{\rho,-\mathbf{k}}=f^\dagger_{\rho,-\mathbf{k}}e^G$, we get that
\begin{equation} \label{eq:f_vs_fdag_on_MF}
    f_{\beta,\mathbf{k}}\ket{\Omega}=(g_\mathbf{k})_{\beta\rho}f^\dagger_{\rho,\mathbf{-k}}\ket{\Omega}.
\end{equation}
Using this in Eq.~\ref{eq:bogoliubov_transform} and forcing $\alpha_{l\mathbf{k}}\ket{\Omega}=0$ gives $(U_\mathbf{k})_{l\sigma}(g_\mathbf{k})_{\sigma\nu}f^\dagger_{\nu,\mathbf{-k}}\ket{\Omega}=(V_\mathbf{k})_{l\sigma}f^\dagger_{\sigma,\mathbf{-k}}\ket{\Omega}$. For this to hold for all $l,\mathbf{k}$ we need simply $U_\mathbf{k}g_\mathbf{k}=V_\mathbf{k}$ for all $\mathbf{k}$, where this is now to be understood as a matrix equation. This is simply the matrix generalisation of the scalar relation $g_\mathbf{k}=v_\mathbf{k}/u_\mathbf{k}$ found in Read-Green \cite{46_Read_Green}.

Following Read and Green further, the mean-field ground state will be a Pfaffian of the fourier transform of $\gk$, as shown in Eq.~\ref{eq:mf_Pfaffian_form} of the main text. To give a concrete example for clarity, for $N=3$ physical bosons with $b=f_1f_2$, so $K=2$ gives
\begin{equation}
\phi(\mathbf{x}_1,\mathbf{x}_2,\mathbf{x}_3)=\text{Pf}\left[\begin{array}{cc|cc|cc}
        0 & g_{12}(\mathbf{x}_1,\mathbf{x}_1) & g_{11}(\mathbf{x}_1,\mathbf{x}_2) & g_{12}(\mathbf{x}_1,\mathbf{x}_2) & g_{11}(\mathbf{x}_1,\mathbf{x}_3) & g_{12}(\mathbf{x}_1,\mathbf{x}_3)\\
        g_{21}(\mathbf{x}_1,\mathbf{x}_1) & 0 & g_{21}(\mathbf{x}_1,\mathbf{x}_2) & g_{22}(\mathbf{x}_1,\mathbf{x}_2) & g_{21}(\mathbf{x}_1,\mathbf{x}_3) & g_{22}(\mathbf{x}_1,\mathbf{x}_3)\\ \hline
        g_{11}(\mathbf{x}_2,\mathbf{x}_1) & g_{12}(\mathbf{x}_2,\mathbf{x}_1) & 0 & g_{12}(\mathbf{x}_2,\mathbf{x}_2) & g_{11}(\mathbf{x}_2,\mathbf{x}_3) & g_{12}(\mathbf{x}_2,\mathbf{x}_3)\\
        g_{21}(\mathbf{x}_2,\mathbf{x}_1) & g_{22}(\mathbf{x}_2,\mathbf{x}_1)  & g_{21}(\mathbf{x}_2,\mathbf{x}_2) & 0 & g_{21}(\mathbf{x}_2,\mathbf{x}_3) & g_{22}(\mathbf{x}_2,\mathbf{x}_3)\\ \hline
        g_{11}(\mathbf{x}_3,\mathbf{x}_1) & g_{12}(\mathbf{x}_3,\mathbf{x}_1)  & g_{11}(\mathbf{x}_3,\mathbf{x}_2)  & g_{12}(\mathbf{x}_3,\mathbf{x}_2) & 0 & g_{12}(\mathbf{x}_3,\mathbf{x}_3)\\
        g_{21}(\mathbf{x}_3,\mathbf{x}_1) & g_{22}(\mathbf{x}_3,\mathbf{x}_1)  & g_{21}(\mathbf{x}_3,\mathbf{x}_2)  & g_{22}(\mathbf{x}_3,\mathbf{x}_2) & g_{21}(\mathbf{x}_3,\mathbf{x}_3) & 0
    \end{array}\right]. \label{eq:actual_wf_example}
\end{equation}
Really, $\gk$ is a $D\times D$ matrix and so is its Fourier transform $\tilde{g}_{\rho\sigma}(\mathbf{r})$. If $\mathbf{x}$ and $\mathbf{y}$ fall on lattice sites $\alpha$ and $\beta$ and the unit cells of $\mathbf{x},\mathbf{y}$ are separated by $\mathbf{r}$, then we have $g_{ab}(\mathbf{x},\mathbf{y})=\tilde{g}_{(\alpha,a),(\beta,b)}(\mathbf{r})$ --- so while we do not make it explicit in Eq.~\ref{eq:actual_wf_example} in favour of notational brevity, $g$ actually respects translational symmetry. The index $(\alpha,a)$ (with $\alpha=1\ldots N_U$ and $a=1\ldots K$) indexes which of the $D=N_UK$ parton sites of the unit cell we consider.

Operationally, we can ignore the background of the $\Vk$ and $\Uk$, simply optimizing the function $g$ in Eq.~\ref{eq:mf_Pfaffian_form} (the algorithm details are described below). We again stress that while we used a mean-field parton Hamiltonian to motivate the form of the ansatz, there are no parton hamiltonians used in our algorithm, only the full bosonic Hamiltonian. But we might still want to recover $\Vk$ and $\Uk$ from the $g$ obtained, for example to compute Chern numbers and mean-field-level pairing. This is the focus of the following section.

\subsection{Recovering mean-field expectations }
The computed $g(\mathbf{x},\mathbf{y})$ may be transformed back into a $\gk$ as described above.
The first step is to relate the VMC-computed $\gk$ back to the  $U_\mathbf{k}$ and $V_\mathbf{k}$ in the Bogoliubov transform Eq.~\ref{eq:bogoliubov_transform}.

We have two constraints on the $\Uk,\Vk$: one comes from the need that $\Uk \gk=\Vk$ and the other is the condition of unitarity on the Bogoliubov transform Eq~\ref{eq:bogoliubov_transform} from $(f_{\sigma,\mathbf{k}},f^\dagger_{\sigma,\mathbf{-k}})$ to $(\alpha_{l,\mathbf{k}},\alpha^\dagger_{l,\mathbf{-k}})$. This second condition reads $\mathcal{U}_\mathbf{k}^\dagger \mathcal{U}_\mathbf{k}=\mathbb{1}$ with
\begin{equation}
    \begin{pmatrix}
        \alpha_{l,\mathbf{k}}\\
        \alpha^\dagger_{l,\mathbf{-k}}
    \end{pmatrix} = \begin{pmatrix}
        \Uk & -\Vk \\
        -V_\mathbf{-k}^* & U_\mathbf{-k}^*
    \end{pmatrix}\begin{pmatrix}
        f_{\sigma,\mathbf{k}} \\ f^\dagger_{\sigma,\mathbf{-k}}
    \end{pmatrix}=\mathcal{U}_\mathbf{k}\begin{pmatrix}
        f_{\sigma,\mathbf{k}} \\ f^\dagger_{\sigma,\mathbf{-k}}
    \end{pmatrix}. \label{eq:full_bogoliubov_trans}
\end{equation}
This is equivalent to the more well-known condition requiring that the Bogoliubovs must obey both $\{ \alpha_{l\mathbf{k}},\alpha^\dagger_{m\mathbf{q}}\}=\delta_{lm}\delta_\mathbf{kq}$ and $\{ \alpha_{l\mathbf{k}},\alpha_{m\mathbf{q}}\}=0$. In what follows, we interpret the $2D\times D$ matrix $\begin{pmatrix} \Uk^\dagger \\ -\Vk^\dagger \end{pmatrix}$ and similar constructions as a set of $D$ vectors with $2D$ components each. Then, the condition that $\Uk\gk=\Vk$, is really just the requirement that $\begin{pmatrix} \Uk^\dagger \\ -\Vk^\dagger \end{pmatrix}$ is orthogonal to $\begin{pmatrix} \gk \\ \mathbb{1}\end{pmatrix}$, ie that each
of the $D$ vectors in one is orthogonal to each of the $D$ in the other. Because $\begin{pmatrix} \gk \\ \mathbb{1}\end{pmatrix}$ is of rank $D$ (guaranteed by the $D\times D$ identity), the dimension of the space orthogonal to it has dimension $2D-D=D$. In fact, it may be checked that a set of $D$ linearly independent vectors orthogonal to it are just the columns of $\begin{pmatrix}  \mathbb{1} \\ -\gk^\dagger \end{pmatrix}$. So all the vectors of $\begin{pmatrix} \Uk^\dagger \\ -\Vk^\dagger \end{pmatrix}$ are within the span of the vectors in $\begin{pmatrix}  \mathbb{1} \\ -\gk^\dagger \end{pmatrix}$. Furthermore, unitarity of Eq.~\ref{eq:full_bogoliubov_trans} necessitates among other things $\Uk^\dagger\Uk+\Vk^\dagger\Vk=\mathbb{1}$, in other words the vectors in $\begin{pmatrix} \Uk^\dagger \\ -\Vk^\dagger \end{pmatrix}$ are orthonormal. This all implies that we may take
\begin{equation}
    \begin{pmatrix} \Uk^\dagger \\ -\Vk^\dagger \end{pmatrix}=\text{orthonorm}\left[ \begin{pmatrix}
    \mathbb{1} \\ -\gk^\dagger \end{pmatrix}\right]. \label{eq:uv_orthonorm}
\end{equation}
We might worry that orthonormalization is not unique and indeed for any $\Uk,\Vk$ satisfying Eq.~\ref{eq:uv_orthonorm}, so do $W\Uk,W\Vk$ for any unitary $W$. This is because the state is defined via $\alpha_{l,\mathbf{k}}\ket{\Omega}=0$, but we could as well have defined it via $\tilde{\alpha}_{l,\mathbf{k}}\ket{\Omega}=0$ with $\tilde{\alpha}_{l,\mathbf{k}}=W_{lr}\alpha_{r,\mathbf{k}}$. Since we do not have a Hamiltonian to distinguish bands, there is no prefered linear combination. All observables must be independent of this arbitrary choice.

By the same logic, but also complex-conjugating and using $\gk^T=-g_\mathbf{-k}$, we may deduce that 
\begin{equation} \label{eq:minus_k_orthogonalization}
    \begin{pmatrix} -V_\mathbf{-k}^T \\ U_\mathbf{-k}^T \end{pmatrix}=\text{orthonorm}\left[ \begin{pmatrix}
    \gk \\ \mathbb{1} \end{pmatrix}\right].
\end{equation}
Because the vectors on the RHS of Eq.~\ref{eq:uv_orthonorm} are orthogonal to those on the RHS of Eq.~\ref{eq:minus_k_orthogonalization}, the same holds for the vectors on the LHS. After hermitian conjugating, this shows that the $D$ rows of $\left(\Uk,-\Vk\right)$ are all orthogonal to the $D$ rows of $\left(-V^*_\mathbf{-k},U^*_\mathbf{-k}\right)$. This and the orthonormalities imply that the combined $2D$ rows of these two vectors form an orthonormal set of vectors which then directly proves that this procedure constructs a set of $\Uk,\Vk$ which make Eq.~\ref{eq:full_bogoliubov_trans} unitary.

There still exists the issue of overall scale. Other than an overall normalization, the wavefunction Eq.~\ref{eq:mf_Pfaffian_form} is independent of the scale of $g$, while both the true mean-field state Eq.~\ref{eq:bdg_gs_form} and also Eq.~\ref{eq:uv_orthonorm} depend non-trivialy on the scale of $g$. This is the case because the mean-field wavefunction is a linear combinations of states with different particle numbers and we are looking only at the component with some fixed number of particles which on its own is insufficient to fix the scale of $g$. The solution is to look at the expectation value of the number of partons at the MF level. Using the transformation Eq.~\ref{eq:full_bogoliubov_trans} and $\alpha_{l,\mathbf{k}}\ket{\Omega}=0$, we find that
\begin{equation} \label{eq:density_K_def}
    n_\mathbf{k}=\sum_{\mu}\frac{\braketmatrix{\Omega}{f_{\mu,\mathbf{k}}^\dagger f_{\mu,\mathbf{k}}}{\Omega}}{\braket{\Omega}{\Omega}} = \text{Tr} \Vk^\dagger \Vk.
\end{equation}
Crucially, this is unchanged if $\Vk\rightarrow W\Vk$ as described above. So really because we do not know the scale of $g$, Eq.~\ref{eq:uv_orthonorm} should be taken as
\begin{equation}
    \begin{pmatrix} \Uk^\dagger(\theta) \\ -\Vk^\dagger(\theta) \end{pmatrix}=\text{orthonorm}\left[ \begin{pmatrix}
    \mathbb{1}\cos(\theta) \\ -\gk^\dagger\sin(\theta) \end{pmatrix}\right]. \label{eq:uv_orthonorm_scaled}
\end{equation}
with $\theta$ determening the overall scale. The idea is then to search over all $\theta$, compute $\Vk(\theta)$ and $N(\theta)=\sum_\mathbf{k}\text{Tr} \Vk^\dagger(\theta)\Vk(\theta)$ and then choose the $\theta^*$ with $N(\theta^*)=NK$ (the desired number of partons at the MF level). The search may be done by bisection on the range $\theta\in[0,\frac\pi2]$. Using the $\Vk$ and $\Uk$ at $\theta^*$, we may then compute correlators and Chern numbers. For example, the mean-field pairing strength can be found to be
\begin{equation} \label{eq:delta_K_def}
    (\Delta_\mathbf{k})_{\nu\mu}=\frac{\braketmatrix{\Omega}{f_{\nu,\mathbf{-k}}f_{\mu,\mathbf{k}}}{\Omega}}{\braket{\Omega}{\Omega}} = (U_\mathbf{k}^\dagger V_\mathbf{k})_{\nu\mu}.
\end{equation}
Again this is independent of the ambiguity of $U_\mathbf{k}\rightarrow W_\mathbf{k} U_\mathbf{k}$, $V_\mathbf{k}\rightarrow W_\mathbf{k}V_\mathbf{k}$ discussed above. Also we have that $\Delta_\mathbf{-k}=-\Delta^T_\mathbf{k}$ which follows when we required column orthogonality in Eq.~\ref{eq:full_bogoliubov_trans}. This is re-assuring and is in fact necessary from the definition Eq.~\ref{eq:delta_K_def} and Fermi anticommutation. Also with a known $U_\mathbf{k}$ and $V_\mathbf{k}$, we may use the fact that the $D$ occupied BdG bands have the form Eq.~\ref{eq:bogoliubov_transform} to compute the Chern number using the non-Abelian algorithm of Ref.~\cite{cherns_alg}. Using the non-abelian algorithm guarantees no dependence on the ambiguity of which exact $\Uk,\Vk$ (or $W\Uk,W\Vk$) is taken.

So given a $g(\mathbf{r})$, we have an algorithm which both figures out the appropriate normalization and then computes the BdG Chern number, pairing strength and can also in principle compute any other correlation function at the mean-field level, as Wick's theorem holds there. Still, we might want to further address the question of why one singles out the particular mean field state where $N(\theta^*)=NK$. After all, the other $\theta$ which represent different parton mean-field states produce the same projected bosonic wavefunction. The main reason is that the mean-field parton wavefunction lies in a larger Hilbert space, much of which is unphysical --- demanding that $N(\theta^*)=NK$ guarantees that as large a component of that wavefunction is actually in the physical subspace as possible, meaning that mean-field values evaluated at $\theta^*$ should be more meaningful for the properties of the physical state --- we again note that one has to be careful when interpreting mean-field values physically.

\subsection{Paired Chern numbers and transitions}
In the unpaired, $\Delta=0$ and uncorrelated limit, the BdG Chern number is just $\chern^\text{Bdg}=2(\chern^\text{P}_1+\chern^\text{P}_2)$. To see why the factor of $2$, we note that the BdG Chern number counts the number of Majorana edge modes, while the usual Chern numbers count the Dirac edge modes. So we expect $\chern^\text{Bdg}_\text{SF}=0$ and $\chern^\text{Bdg}_\text{FCI}=4$ in the unpaired limit. Assuming a small pairing $\Delta$ does not close any gaps, this should also hold at mean-field for the paired states.
We still have the projective translation symmetry for partons $T_1^\text{P}T_2^\text{P}=-T_2^\text{P}T_1^\text{P}$ which means a doubling in the spectrum, implying that the Chern number jumps by two units, or really double the jump that would happen otherwise.  Furthermore, the BdG nature of the problem dictates that a band gap closure at a point $\mathbf{k}$ necessitates another closure at $-\mathbf{k}$. Following logic similar to \cite{17_PhysRevLett.115.026802}, we may then argue that a generic band-gap closure occurs at four points in the BZ, $\mathbf{k},\mathbf{k}+(0,\pi),\mathbf{-k}$ and $\mathbf{-k}+(0,\pi)$. This leads to the conclusion that the jump from $\chern^\text{Bdg}=4$ to $\chern^\text{Bdg}=0$ is natural and that the SF-FCI transition may occur without fine-tuning.

But the question remains of what happens if $\mathbf{k}=(0,\frac\pi2)$ or $\mathbf{k}=(0,0)$ in which case two pairs of the points coincide --- so that the projective translation doubling and the BdG doubling in fact produce the same doubles. Is it possible that we only get a change of $\chern^\text{BdG}$ by two units? In the case $\mathbf{k}=(0,\frac\pi2)$, this would come in the form of a realness condition, ie if the gapless fermions associated with the gap closures are denoted by $\psi_1$ and $\psi_2$, we would have $\psi_1^\dagger=\psi_2$ (and $\psi_{1,2}$ transform into each other under the action of $T_1^\text{P}$). In the case $\mathbf{k}=0$, this simply means that the spectrum-doubling happens in terms of Majorana modes instead of in terms of Dirac particles. We again note that while we have not found arguments prohibiting this phase, we have also not found a bosonic Hamiltonian realizing it. In any case, having this transition in terms of only Dirac fermions is impossible --- so a genuine BdG Hamiltonian is needed. What this means is that solely $\MFEV{f_1^\dagger f_1}$, $\MFEV{f_2^\dagger f_2}$ and $\MFEV{f_1f_2}$ being nonzero is not enough --- we need either $\MFEV{f_1^\dagger f_2}\neq0$ or $\MFEV{f_1f_1}\neq 0$.

Crucially, near the FCI-SF transition observed in Fig.~\ref{fig:checkerboard_ed_comparison} of the main text we find that only $\MFEV{f_1^\dagger f_1}$, $\MFEV{f_2^\dagger f_2}$ and $\MFEV{f_1f_2}$ are nonzero. A simple particle-hole conjugation of $f_2$ would get rid of all anomalous terms, meaning that no Majorana modes can be present. (Such a transformation is analogous to the transformation that takes a BCS superconductor to an excitonic superfluid.)
This is the basis for our claim in the main text that this $\chern^\text{BdG}=2$ phase is unlikely to be an intermediary between FCI and SF at the transition observed. But we do not see any reason why this phase could not exist elsewhere in parameter space.

One way this $\chern^\text{BdG}=2$ phase could be realized is with parton $f_2$ in a $\chern=1$ (non-BdG) band, while $f_1$ transitions from a $\chern^\text{P}=1$ (non-BdG) band into a $\chern=0$ (BdG) band via a double-Majorana gap closure as discussed above. Following Barkeshli and McGreevy \cite{05_barkMcGreevyTheory}, we characterize this phase by its electromagnetic response. The parton $f_1$ feels a gauge field $\frac12A_\mu+a_\mu$ while $f_2$ feels $\frac12A_\mu-a_\mu$ (here $A_\mu$ is external electromag field and $a_\mu$ is the emergent $U(1)$ ``gluing'' gauge field from the parton construction). Since $f_1$ is in a paired state, the gauge field it feels, $b_\mu=\frac12A_\mu+a_\mu$ gets Higgsed, ie we get a term $c b_\mu b^\mu \subset \mathcal{L}$. Integrating out the other parton, $f_2$ gives the simple IQHE response $\frac{1}{4\pi}\csterm{(\frac12A-a)}{( \frac12A-a)} \subset \mathcal{L}$. Now when integrating out $a_\mu$, the most relevant at large distance is the Higgs term which simply sets $b_\mu\rightarrow0$ meaning that we get $a_\mu\rightarrow-\frac12A_\mu$ and so the final electromagnetic response is simply $\mathcal{L}=\frac{1}{4\pi}\csterm{A}{A}$, so the system is an insulator with a unique ground state but with an IQHE response. Note that this looks somewhat similar to some of the discussion of \cite{15_Senthil_theory}, which sees a transition from a $\nu=\frac23$ fermionic FCI either to a CDW insulator with no Hall response or one with an IQHE response. We leave more study of this phase to further work.

We should note that Barkeshli and McGreevy \cite{05_barkMcGreevyTheory} similarly discussed a phase ``in between'' FCI and SF, theirs with (non-BdG) parton Chern numbers $(\chern^\text{P}_1,\chern^\text{P}_2)=(1,0)$ which turns out to be a CDW Mott insulator as discussed above --- one has to break translation symmetry to have the Chern number jump by one unit. Our phase is fundamentally different since it requires parton pairing, which this Mott insulator does not.

\subsection{BdG many-body Chern number}
It can be shown that the appropriately normalized BdG mean-field state is (prime on the product means multiply over each ${\bf k},-{\bf k}$ pair only once)
\begin{equation} \label{eq:normalized_bcs_mf}
    \ket{\Omega}=\prod_\mathbf{k}'\det \Uk e^{(\gk)_{\sigma\rho}f_{\sigma,\mathbf{k}}^\dagger f^\dagger_{\rho,\mathbf{-k}}}\ket{\emptyset}.
\end{equation}

Now say we twist the boundary conditions by $\theta_1$ and $\theta_2$ in the two directions and say we look at the holonomy of $\ket{\Omega}$ as we drag it around a closed loop $(\theta_1(t),\theta_2(t))$. The holonomy will come from the holonomy of the $\alpha's$ in Eq.~\ref{eq:bogoliubov_transform}. Each $\alpha_{l\mathbf{k}}$ gets dragged around a loop centered around $\mathbf{k}$, following $\mathbf{k}(t)=\mathbf{k}+\left(\theta_1(t)/KN_x,\theta_2(t)/KN_y\right)$. Here, $N_x$ and $N_y$ are the system dimensions and $K$ is the number of partons: for $K$ partons, we only need to apply $\theta/K$ of a twist to each in order to get a full WF twist of $\theta$. It is not a priori obvious that the twist should be equal for both partons --- a possible justification is the fact that they see equal charge. But we note that this procedure recovers the correct many-body Chern number in our case.

For a small loop (any loop in $\theta$-space will inmply small loops in $\bf k$ space in the thermodynamic limit $N_x,N_y\rightarrow\infty$) we will have after the holonomy $\alpha_{l\mathbf{k}}\rightarrow W_{\mathbf{k},lm}\alpha_{m\mathbf{k}}$ (it is assumed that there is a finite gap to the $\alpha^\dagger_{l,\mathbf{-k}}$  states at positive energy). For a loop area $A_\mathbf{k}$, we have $\det W_\mathbf{k}=e^{iA_\mathbf{k}\mathcal{F}_\mathbf{k}}$ where $\mathcal{F}_\mathbf{k}$ is the (trace of the) total Berry curvature of the filled bands.

The effect of $\alpha_{l\mathbf{k}}\rightarrow W_{lm}\alpha_{m\mathbf{k}}$ is really that our matrices transform as $\Uk\rightarrow W_\mathbf{k}\Uk$ and $\Vk\rightarrow W_\mathbf{k}\Vk$. Note that this leaves $\gk$ invariant, but crucially introduces a factor of $\prod_\mathbf{k}'\det W_\mathbf{k}=\exp\left[i\sum_\mathbf{k}' A_\mathbf{k} \mathcal{F}(\mathbf{k})\right]$ to Eq.\ref{eq:normalized_bcs_mf}. Now using the fact that $A_\mathbf{k}=\frac{1}{K^2N_xN_y} A_\mathbf{\theta}$ (with $A_\mathbf{\theta}$ the area of the $\theta$-space loop), we get that the holonomy of $\ket{\Omega}$ over a closed loop is $\exp\left[i \frac{A_\mathbf{\theta}}{K^2} \frac{1}{N_xN_y}\sum_\mathbf{k}' \mathcal{F}(\mathbf{k})\right]$. Now we use the fact that (following from the BdG nature of the problem) $\mathcal{F}(\mathbf{-k})=\mathcal{F}(\mathbf{k})$ to extend the sum to all $\mathbf{k}$ and take the thermodynamic limit to change it to an integral.
Thus the holonomy of $\ket{\Omega}$ after a loop of area $A_\mathbf{\theta}$ is $e^{i A_\mathbf{\theta} \mathcal{F}_\theta}$ where
\begin{equation}
    \mathcal{F}_\theta = \frac{1}{2K^2}\int\text{d}^2\mathbf{k}~\mathcal{F}(\mathbf{k})=\frac{1}{2K^2} \frac{\chern_\text{BdG}}{2\pi}
\end{equation}
is the many-body berry curvature. The integral of this quantity is the many-body Chern number which leads us to the relation \begin{equation}
    \chern_\text{MB}=\frac{1}{2K^2}\chern_\text{BdG}
\end{equation}
This correctly gives $\chern_\text{MB}=\frac12$ for the FCI and $\chern_\text{MB}=0$ for the SF phases.

\section{Self-consistent mean-field \label{sup:SCMF}}
Our treatment so far was to use the parton picture to motivate the wavefunction ansatz Eq.~\ref{eq:projected_bose_WF}, which we then optimized in VMC. But can we trust the parton picture even more, for example only evaluating the Hamiltonian at the mean-field level, $\MFEV{H^\text{B}}$ and minimizing that? While less accurate, such a procedure does not require Monte Carlo sampling and would thus give access to much larger system sizes. An immediate problem is the bosonic number operator $\hat{n}_\text{B}=b^\dagger b$. In the projected subspace, $\hat{n}_\text{B}=\hat{n}_1=\hat{n}_2$ holds as an operator identity, but at the mean-field level, $\MFEV{\hat{n}_\text{B}}=\MFEV{\hat{n}_1}\MFEV{\hat{n}_2}+|\Delta|^2$ with $\Delta=\MFEV{f_1f_2}$ (the mean-field state should be chosen so that $\langle f_2^\dagger f_1\rangle_\text{MF}=0$ \cite{wen_book}). If we fill one of two bands at $\nu=\frac12$, we expect $\MFEV{\hat{n}_{1,2}}=\frac14$. In the FCI phase with no pairing, $\MFEV{\hat{n}_\text{B}}=\frac{1}{16}$, which means we get a quarter as many bosons as we wanted. This is because we only see a boson when two partons happen to hop on the same physical site, which simply does not happen often enough. For a similar reason, all boson hopping expectations are generally roughly a quarter the size. The problem then is that the expectation $\MFEV{H^\text{B}}$ will often be lowest when $\MFEV{\hat{n}_B}$ is large, thus when $|\Delta|$ is big which leads to this SCMF procedure favouring a paired-parton state even where more sophisticated methods find the FCI. A simple remedy such as minimizing $\MFEV{H^\text{B}}/\MFEV{\hat{n}_\text{B}}$ incorrectly displays a first-order transition. Instead, we repeat this SCMF scheme for all values of $|\Delta|$ along the transition. For the checkerboard model, we find that this SCMF method matches the transition point of the VMC when $|\Delta|\sim0.4$. At points near the transition, we are able to look at the Berry curvature of the parton BdG bands, as shown in Fig.~\ref{fig:scmf_berry}.
\begin{figure}
    \centering
    \includegraphics[width=3.5in]{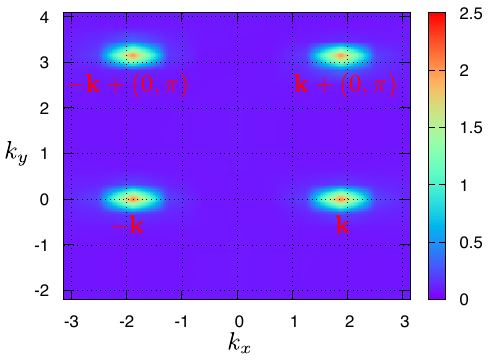}
    \caption{Parton BdG Berry curvature obtained from self-consistent mean-field with $|\MFEV{f_1f_2}|=0.4$ enforced. The system here is in the FCI phase, but close to the SF transition. We see four points where the curvature is concentrated --- these are the points where a band gap will close. As discussed in the text, we expect four such points to occur simultaneously. Note that $\mathbf{k}=(k_x,k_y)$ refers to momenta in the reduced, partonic BZ.}
    \label{fig:scmf_berry}
\end{figure}
We see four points where the Berry curvature is concentrated, corresponding to the four band gap closures. Denoting the momentum of one by $\mathbf{k}$, the BdG nature of the Hamiltonian leads to a closure at $-\mathbf{k}$. Furthermore, as discussed above, projective translational symmetry requires gap closures at $\mathbf{k}+(0,\pi)$ and $-\mathbf{k}+(0,\pi)$. The closures seen in Fig~\ref{fig:scmf_berry} are consistent with this, confirming that the transition seen is of this projective-translation-protected gap closure type. A similar construction of four gap closures was found in a related scenario in \cite{17_PhysRevLett.115.026802}.

\end{document}